\documentclass[12pt,a4paper,twoside]{article}        
\usepackage[english]{babel}                          
\usepackage[cp1251]{inputenc}                        
\usepackage{mmgart-dvips}                            
\usepackage{cite}                                    

\begin{document}

\setcounter{page}{29}                                
\thispagestyle{empty}                                
\begin{heading}                                      
{Volume\;3,\, N{o}\;3,\, p.\,29 -- 59\, (2015)}      
{}
\end{heading}                                        

\begin{Title}
Computer simulation of electronic excitations in beryllium
\end{Title}

\begin{center}
\Author{}{A.V. Popov}
\end{center}


\begin{flushleft}
\Address{}{\!\!\!Department of Special Modern Materials, Polzunov Altai State Technical University, Lenin prospect, 46, Barnaul, Russian Federation, 656038}


\Email{Popov.Barnaul@mail.ru}
\end{flushleft}

\Headers{A.V. Popov}{Computer simulation of electronic excitations in beryllium}

\begin{flushleft}                                 
\small\it Received 9 December 2015, in final form 
29 December. Published 31 December 2015.          
\end{flushleft}                                   


\Thanks{\mbox{}\\
\copyright\,The author(s) 2015. \ Published by Tver State University, Tver, Russia}
\renewcommand{\thefootnote}{\arabic{footnote}}
\setcounter{footnote}{0}

\Abstract{An effective method for the quantitative description of the electronic excited states of polyatomic systems is developed by using computer technology. The proposed method allows calculating various properties of matter at the atomic level within the uniform scheme. A special attention is paid to the description of beryllium
atoms interactions with the external fields, comparable by power to the fields in atoms, molecules and clusters.}



\newpage                               
\renewcommand{\baselinestretch}{1.1}   


\section{Introduction}

By matter one usually means a set of electrons and nuclei which constitute atoms, molecules and clusters of a gas, liquid, plasma or solid state. For the interaction describing of this set of particles with an external field it is necessary to consider not only the influence of the field on matter, but also the matter influence on
the field. It is especially important to consider this mutual influence for the fields comparable to those in atoms, molecules, or clusters. Such fields became quite achievable with the advent of powerful lasers. The impact of powerful lasers on matter leads to reorganization of electron shells and electronic structure of matter. Such a structure reorganization can also result in chemical reactions. Note, that the field of mutual influence of atoms or molecules is rather wide.

Small metal clusters and metal nanoparticles have received a considerable deal of experimental and theoretical attention in recent years \cite{Jena,Boon,Schmid}. The study of molecular clusters is also of great interest \cite{Shridhar}. Nevertheless, there is still no complete understanding on how the properties of a cluster evolve with the size and at what size a particular metal cluster
exhibits the properties of the bulk metal.

Beryllium, being one of the simplest elements, became the benchmark for computational \emph{ab initio} methods, with hundreds of publications. Most of the theoretical works have been aimed at elucidating the ground state properties of small beryllium clusters. Nevertheless, the bonding nature of beryllium aggregates is not fully understood even for the smallest $Be_{2}$ cluster.

This paper will provide a brief overview of the main experimental and theoretical works that have been carried out for small beryllium clusters and then an overview of theoretical description of strong electronic excitations in beryllium.


\section{Bonding in small beryllium clusters}

The beryllium atom has four electrons, so it can be adequately described by nonrelativistic Hamiltonian. Beryllium clusters are suitable for high-level coupled cluster methods with large basis sets. Experiments have been limited to the $Be_{2}$ dimer \cite{Merritt}.
Spectroscopic observations show that the $Be_{2}$ molecule is weakly bound, with a bond distance of 2.45 {\AA}, surprisingly stronger and shorter than those between other similar sized close-shell atoms \cite{Schaefer,McLaughlin}. The more strongly bound $Be^{+}_{2}$ molecule has also generated experimental interest
\cite{Merritt_Phys_Chem_Chem_Phys,Antonov}, as have small oxides of $Be$ such as $BeOBe$ \cite{Merritt_J_Phys_Chem_A} and diatomic beryllium carbide $BeC$ \cite{Barker}.

Note, that the first empirical confirmation of $Be_{2}$ existence was reported in the 1980s
\cite{Bondybey_J_Chem_Phys,Bondybey_Chem_Phys_Lett,Bondybey_Science}.
The well depth for the ground electronic state was reported to be $D_{e}=790±30 cm^{-1}$. This result was not accurate and the true error is much larger than the estimated error bars. In 2009 a new experiment was performed by Merritt \emph{et al.} \cite{Merritt} and the interaction energy was found to be $929.7±2.0$ $cm^{-1}$. The study also increased the number of identified vibrational levels from 5 to 11. The authors suspected that one more level could exist,
but their empirical fit to the measured data predicted only 11 levels, with the vibrational quantum number $v$ ranging from 0 to 10. The existence of a twelfth level, just 0.44 $cm^{-1}$ below the dissociation limit, was confirmed by using the “morphed” theoretical potential energy curve \cite{Patkowski}. But at first even the existence of the beryllium dimer was in question. In 1929, when
Herzberg \cite{Herzberg} failed to produce $Be_{2}$, he concluded that the interaction between two ground-state beryllium atoms is repulsive.

The first theoretical investigations, which include valence bond \cite{Bartlett}, self-consistent field \cite{Fraga_1961,Fraga_1962}, and configuration interaction \cite{Bender} calculations, also showed that the $Be_{2}$ dimer potential is repulsive. In the 1980s Liu \emph{et al.} \cite{Liu,Lengsfield} showed that $Be_{2}$ has a
deep minimum (806 $cm^{-1}$) at a short bond length (2.49 {\AA}), while several theoretical studies \cite{Blombeg,Harrison_1983,Harrison_1986} concluded that the $Be_{2}$ potential has a shallow van der Waals minimum at a large internuclear distance (4.5 to 5.25 {\AA}). Later publications confirm the existence of $Be_{2}$ van der Waals molecule at long-range distance \cite{Klopper,Watts}. On the whole the key results of calculations before 1995 are cited in \cite{Roeggen}. More recent publications can be found in Ref. \cite{Lesiuk}. Potential energy curves for the ground state of $Be_{2}$ were calculated using a variety of theoretical methods: self-consistent field (SCF), second-order Moller-Plesset perturbation theory (MP2), coupled cluster with single and double excitations (CCSD), multireference configuration interaction (MRCI), density functional theory (DFT) and their modifications. The MRCI potential with the Davidson correction is the closest to the experimental result in terms of both depth and shape \cite{Heaven_2011}.

As for the $Be_{3}$ trimer, there are still no spectroscopic data, nevertheless it has been the subject of many theoretical studies. On the whole, $Be_{3}$ shows a similar behavior to that observed in the dimer. Like in that case, the $Be_{3}$ cluster is not bound at the
SCF level \cite{Harrison_1986,Lee}. CCSD calculations yield a shallow binding minimum, while MRCI calculations lead to a well-defined minimum at distances shorter than those observed in CCSD calculations \cite{Harrison_1986,Lee,Watts}. The DFT B3LYP binding
energy almost doubles the MRCI value \cite{Beyer,Srinivas}. Note that the full configuration interaction (FCI) study of the ground state of the neutral beryllium trimer shows a secondary van der Waals minimum at a long bond distance \cite{Junquera-Hernandez}. The dissociation energy $D_{e}$ of 17.2 kcal/mol corresponds to this value. This $D_{e}$ can be compared with the MRCI result of Watts \emph{et al}. \cite{Watts} who reported 13.9 kcal/mol for the
equilibrium distance of 2.32 {\AA}. Harrison and Handy
\cite{Harrison_1986} reported a value of 19 kcal/ mol for $D_{e}$, assuming a minimum at 2.23 {\AA}. Rendell \emph{et al.} \cite{Rendell} reported a MRCI value of 22.5 kcal/mol and the equilibrium distance of 2.22 {\AA}. Some calculations involving greater basis sets are also available (see Table~\ref{tab1}). On the whole, the binding in the $Be_{3}$ trimer has a mixed physical (van der Waals) and chemical (nonadditive exchange) nature \cite{Kaplan_2000}.

$Be_{4}$ is the smallest beryllium cluster that is stable at the SCF level of theory \cite{Bauschlicher,Brewington,Lee}. \emph{Ab initio}
calculations predict consistently a tetrahedral ($T_{d}$) minimum energy structure. The bond length of the $Be_{4}$ tetramer is estimated to be 2.07 {\AA} and the dissociation energy is about 31 $cm^{-1}$. The analysis of the many-body forces for $Be_{4}$ indicate that the three-body interaction contribution to the bonding is dominant \cite{Blaisten-Barojas,Ascik_2011}.

The most stable structure of $Be_{5}$ is a trigonal bipyramid ($D_{3h}$) in the singlet ground state. Different isomers of the $Be_{5}$ have been studied in \cite{Zhao_2014}. Note, that for the $Be_{5}$ clusters, no results of quality beyond CCSD(T) are available, so the role of the higher-order terms cannot be directly assessed. Taking into account the trends for $Be_{2-4}$ clusters, it
is possible to conclude that by increasing the size of the cluster, CCSD(T) becomes more adequate \cite{Sulka2012}.

As for the $Be_{6}$ cluster, several $Be_{6}$ ground state structures were suggested in the literature~\cite{Merritt,Lee}, differing both in the shape and multiplicity. For example, Srinivas and Jellinek~\cite{Srinivas} suggest the quintet ($O_{h}$) state to be the most stable, while Beyer \emph{et al.} \cite{Beyer} prefer the triplet ($C_{2v}$) ground state. Sulka \emph{et al.} confirm that the ground state structure of $Be_{6}$ is rather sensitive to the applied
computational method and basis set, although the differences may not always be large \cite{Sulka2012}.

Table~\ref{tab1} presents the results of the structures and dissociation energies obtained by different authors for small beryllium clusters.

The equilibrium geometries of the ground-state and lowlying isomers of $Be$ clusters containing up to 41 atoms can be found in~\cite{Cerowski}.

\begin{table}[!hB]
\caption{Bond lengths ($r$, {\AA}) and dissociation energies
($D_{e}$, kcal/mol) for small beryllium clusters $Be_{n}$ (n=3-6).}
\label{tab1}
\begin{center}
\begin{tabular}{|c|c|c|c|c|c|c|c|c|c|c|c|}
\hline
  & \cite{Lee} & \cite{Beyer} & \cite{Srinivas} & \cite{Junquera-Hernandez}
  & \cite{Ascik_2011} & \cite{Sulka2012} & \cite{Ascik_2012} & \cite{klopper_1993}
  & \cite{lee_2005} & \cite{sudhakar_1993}& \cite{Wang} \\
\hline
\multicolumn{12}{|c|}{$Be_{3}$} \\
\hline $r$      &2.23 & 2.48 & 2.20 & 15.00 & - & 2.22 & - & 2.22 & 2.23 & 2.24 & 2.18 \\
\hline $D_{e}$  &22.4 & 32.2 & 40.8 & 17.2  & - & 25.3 & - & 26.9 & 24.8 & 21.6 & 33.2\\
\hline
\multicolumn{12}{|c|}{$Be_{4}$} \\
\hline $r$      &2.06 & 2.03 & 2.07  & - & 2.04 & 2.07 & - & 2.06 & 2.06 & 2.09 & 2.05 \\
\hline $D_{e}$  &77.3 & 97.4 & 110.7 & - & 89.7 & 31.0 & - & 87.9 & 89.7 & 83.4 & 95.0 \\
\hline
\multicolumn{12}{|c|}{$Be_{5}$} \\
\hline $r$      &2.02 & 1.99  & 2.02  & - & - & 2.02  & 2.00  & - & - & 2.03  & 2.01  \\
                &2.08 & 2.03  & 2.09  & - &   & 2.08  & 2.06  &   &   & 2.08  & 2.08  \\
\hline $D_{e}$  &110  & 134.3 & 148.7 & - & - & 128.1 & 197.5 & - & - & 122.1 & 131.4 \\
\hline
\multicolumn{12}{|c|}{$Be_{6}$} \\
\hline $r$      &- & 1.86  & 2.07  & -  & -  & 1.89  & - & - & - & 1.87  & 1.88 \\
                &  & 2.96  &       &    &    & 2.47  &   &   &   & 2.03  & 2.05  \\
\hline $D_{e}$  &- & 165.2 & 178.5 & -  & -  & 157.1 & - & - & - & 136.6 & 131.4 \\
\hline
\end{tabular}
\end{center}
\end{table}

\section{Beryllium atoms in intense fields}

\subsection{Standard computational methods}\label{3.1}

For full and deep understanding of what happens during the formation of matter such a theory is needed that allows calculating various properties of matter within the uniform scheme. At the same time a special attention should be paid to such methods which are based on the first principles of quantum physics. Calculations are also important because many parameters are easier to compute than to measure directly in the experiment. Moreover it is possible even now to simulate new properties of matter by using theoretical data based on a good theory. Note that the formation of matter is influenced by a large number of competing factors, and therefore it is impossible
to be limited only by qualitative considerations, the exact quantitative calculations are needed. At the same time the choice of a calculation method, its physical correctness and mathematical accuracy are of great importance. A certain progress has been made in the theoretical study of the electronic structure of matter in recent years. Methods of calculating the electronic structure of matter in the ground state are the most popular now. However, the
thermal, vacancy, impurity, electromagnetic and other excitations in which the electrons of real matter reside are of greater interest. Besides, the measurement of any characteristics of an electron in the ground state leads to an inevitable impact on it which automatically transfers the particle to the excited state.

One way to describe the electronic excitation is to use many-body perturbation theory that reconciles all interactions between particles. This theory works well when perturbations are small. The model of a free electron gas with interaction between the electrons (DFEG), also frequently called the Fermi liquid theory, is one of
the simplest models of the many-body solid state theory. The basic results of the DFEG theory are obtained in the classic works of Quinn and Ferrell \cite{Quinn} and Ritchie \cite{Ritchie}, who first studied electron dynamics using the self-energy formalism of the many-body theory. In this formalism the damping rate (inverse lifetime) of an excited state is determined by the imaginary part of the electron self-energy. Within this formalism the evolution of a
system is described by a single-particle Green's function. An exact set of equations, for which, in principle, one can obtain a self-consistent solution for the Green's function, was obtained by Hedin \cite{Hedin}. However the Hedin equations are extremely complicated, so their exact solution is impossible even in the model of a free electron gas.

GW-method is one of the simplest methods for programming and for practical calculations. However, the first iteration step, leading to the GW-approximation and its modifications, does not provide reliable description of the electronic structure for many materials. The second iteration step usually leads to even less agreement between the calculation results and the experimental data. Labor-consuming implementation of these approaches is the major constraint for their wide application.

The Hohenberg-Kohn theorem \cite{Hohenberg} made it possible to prove that the density functional theory provides an accurate description of the properties of a system of interacting particles in the ground state. Simple computational methods of the electronic structure for different materials were developed based on this theory. The main difficulty arising here is that the form of the
exact functional isn't known \cite{Sarry}. Modern attempts to construct such \emph{ab initio} functional do not give the expected results \cite{Ipatov}. Multiconfiguration approach gives the results which describe the experimental data much better. However, the calculations made within this extremely weakly converging approximation are so time-consuming that they are used only for systems with a small number of electrons.

A novel method, which helps to overcome these difficulties, is presented in the next section.

\subsection{The idea of the novel method}

Let us consider the spectral problem for the ground state of a
many-electron atom

\begin{equation}
\label{1} \textbf{H}\psi_{i}(\textbf{r})=E\psi_{i}(\textbf{r})
\end{equation}
Hereafter we will omit the index $i$, which denote the $i$-th state
of the electron in the atom, because the Schr\"{o}dinger equation
has the same form for all states.

It is known that the standard solution of Eq. \ref{1} assumes a
separation of the radial and angular variables:
\begin{equation}
\label{2}
\psi=R(r)Y(\theta, \varphi),
\end{equation}
and the angular function satisfies
\begin{equation}
\label{3}
\left(\frac{1}{sin\theta}\frac{\partial}{\partial\theta}\left(sin\theta\frac{\partial}{\partial\theta}\right)+
\frac{1}{sin^{2}\theta}\frac{\partial^{2}}{\partial\varphi^{2}}\right)Y(\theta,\varphi)=
-l(l+1)Y(\theta,\varphi).
\end{equation}

If the requirement is imposed that the solutions of Eqs. \ref{3} be regular on a
sphere for $0\leq\theta\leq\pi$ and $0\leq\varphi\leq2\pi$, continuous at $\theta=0$ and $\theta=\pi$, and satisfy the condition $Y(\theta,\varphi+2\pi)=Y(\theta,\varphi)$, we arrive at an eigenvalue problem that admits solutions only with the integer values of $l=0, 1, 2, ...$ and $m=0,\pm1,\pm2,...,\pm l$.

While describing excitations it is necessary to consider that excitations can spontaneously decay during a finite time. So, the wave function which describes this system should contain such an exponential factor that all probabilities determined by the square of the modulus of the wave function should decay:
\begin{equation}
\label{4}
\Psi(\textbf{r},t)=\psi(r)exp\{-i(E-i\Gamma/2)t/\hbar\}.
\end{equation}
The function $\psi(r)$ on the right side of Eq. \ref{4} can be
conveniently sought in the form of  Eq. \ref{2}. Moreover, let the
angular function also satisfy Eqs. \ref{3}. Since
$Y(\theta,\varphi)$ is a singlevalued function of $\varphi$, we can
restrict ourselves to the integer values of the number $m$. However,
having “allowed” the orbital transitions, we have to abandon the
restrictions imposed on $l$ in Eqs. \ref{3}, considering it as a
complex number $L=l+x+iy$, where we assume $l$ to be an integer as
above and $|x|<1$ and $|y|<1$.

Note that the idea of considering the complex angular momentum is actively used by more than 50 years in scattering theory, the so-called Regge theory, the study of the analytic properties of scattering as a function of angular momentum, where the angular momentum is not restricted to be an integer but is allowed to take
any complex value~\cite{Regge}. Regge theory has a number of advantages, the main one is a sharp reduction in the number of degrees of freedom necessary to consider the quantum mechanical scattering.

\begin{figure}[!hB]
\center{\includegraphics[width=1\linewidth]{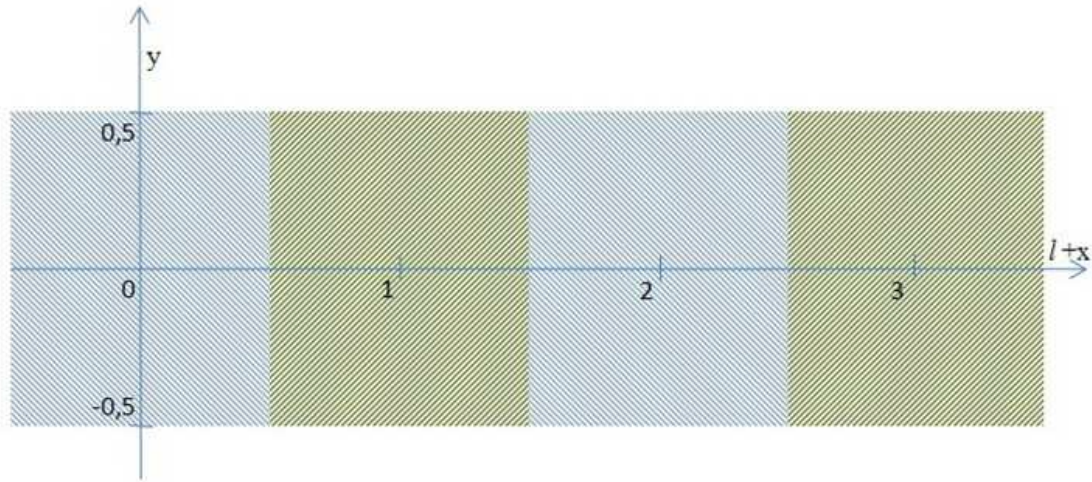}}
\caption{The range of variation of a complex number $L = l + x +
iy$.} \label{figure1}
\end{figure}

The above proposed approach allows us to reduce the problem of seeking the energy spectrum of the excited states of an atom to the eigenvalue problem of the stationary Schr\"{o}dinger equation with complex additives~\cite{Popov2002}:

\begin{equation}
\label{5}
\left(\textbf{H}+\frac{\hbar^{2}}{2m}\frac{u+iv}{r^{2}}\right)\psi(\textbf{r})=\varepsilon\psi(\textbf{r}).
\end{equation}
Here $\textbf{H}$ is the energy operator for an atom in the ground state, $u=x(x+2l+1)-y^{2},  v=y(2x+2l+1)$. Note that, for $y\neq0$ Eq. \ref{5} is essentially non-Hermitian, with complex values of energy $\varepsilon=E-i\Gamma/2$. By searching through all possible values $|x|< 0.5, |y| < 0.5$ (see
Figure~\ref{figure1}) in seeking self-consistent solutions to Eq.~\ref{5}, we can determine the spectral parameters of an excited atom from the minimum value of the real part of its total energy. In particular, based on the uncertainty relation $\Gamma\sim\hbar/t$, we can estimate the decay time of the calculated excited state with any energy equal to the real part of the total energy of an atom.

It should be noted that non-Hermitian quantum mechanics is
increasingly used \cite{Moiseyev}. For example, in quantum optics, where the complex index of refraction are used; in quantum field theory, where the parity–time symmetry properties of the Hamiltonian are investigated, and even in those cases where a formalism of quantum mechanics is applied to the description of the problems associated with classical statistical mechanics or diffusion in biological systems. In some cases, the introduction of complex potentials can greatly simplify the numerical calculations and help one to avoid the interference effects caused by the reflection of the propagating wave packet from the edge of the grid.

As for the above described idea of considering the width of the energy levels, it was first used for finding the wave functions of excited hydrogen atoms \cite{Janavicius}. However, the problem has not been finished in this paper. In Ref.~\cite{Popov2012} we have shown the result of the solution of the spectral problem for the hydrogen atom within the formalism proposed above.

\subsection{Energy spectrum of the beryllium atom}

Hereafter, we will use the system of atomic units. In order to change over to this system, it is sufficient to consider that, in all the relations used, the Plank constant is $\hbar = 1$, the square of the elementary charge is $e^{2} = 2$, and the electron mass is $m = 1/2$. In this case, the energy is measured in Rydbergs and the distance is expressed in bohr. For energy, 1 $Ry$ = $1.09737
\times 10^{5} cm^{-1}$; for distance, 1 bohr = 0.52917 ${\AA}$.

Table~\ref{tab2} shows the results of calculations of the total
energy $E_{tot}$, the energy  $E_{1s}$ and $E_{2s}$ of electrons in
the $1s$, $2s$ states, respectively, obtained using the Hartree-Fock
approximation for the beryllium atom. The results of calculations
using just 9 basis functions of $s$-symmetry, already give good
agreement with the results of other calculations. Adding to the
basis functions of $2p$-symmetry leads to a slight decrease in the
total energy. Nevertheless, it should be noted that the experimental
value of the total energy -29.338665 $Ry$, specified in
\cite{Lindroth}, lies slightly below the calculated values. This is
probably due to the fact that all the calculations were made for the
isolated atom. While achieving the complete isolation of the atom is
impossible during the experiment. That leads to the emergence of the
additional motions, such as the rotational motion of the atom during
the collision of atoms in the gas, and oscillating motion of the
electron shells over the nucleus.

\begin{table}[!hB]
\caption{Results of calculations of the total energy $E_{tot}$, the energy  $E_{1s}$ and $E_{2s}$ of electrons in the $1s$, $2s$ states, respectively, made by the Hartree-Fock approximation for the beryllium atom.} \label{tab2}
\begin{center}
\begin{tabular}{|c|c|c|c|c|c|}
\hline
  & \cite{Clementi1974} & \cite{Huzinaga} & \cite{Bunge} & $s$& $sp$\\
\hline
$E_{tot}, Ry$ & -29.146042 & -29.145158 & -29.146046 & -29.145719 & -29.146106 \\
\hline
$E_{1s}, Ry$ & -9.46534 & -9.46446 & -9.46534 & -9.46458 & -9.46492 \\
\hline
$E_{2s}, Ry$ & -0.61054 & -0.61838 & -0.61854 & -0.61736 & -0.61891 \\
\hline
\end{tabular}
\end{center}
\end{table}

Problem (\ref{5}) was solved numerically in the basis set of Gaussian functions using nine, six, and three functions in the expansions with respect to $l = 0, 1,$ and $2$, accordingly. Our estimates demonstrated that the basis of such a length is quite suitable for the solution of Eqs.\ref{5} by the Roothaan method for a beryllium atom if $|x| < 0.03$ and $|y| < 0.03$ \cite{Popov2005}.

\begin{figure}[!hB]
\begin{minipage}[!hB]{0.45\linewidth}
\center{\includegraphics[width=1\linewidth]{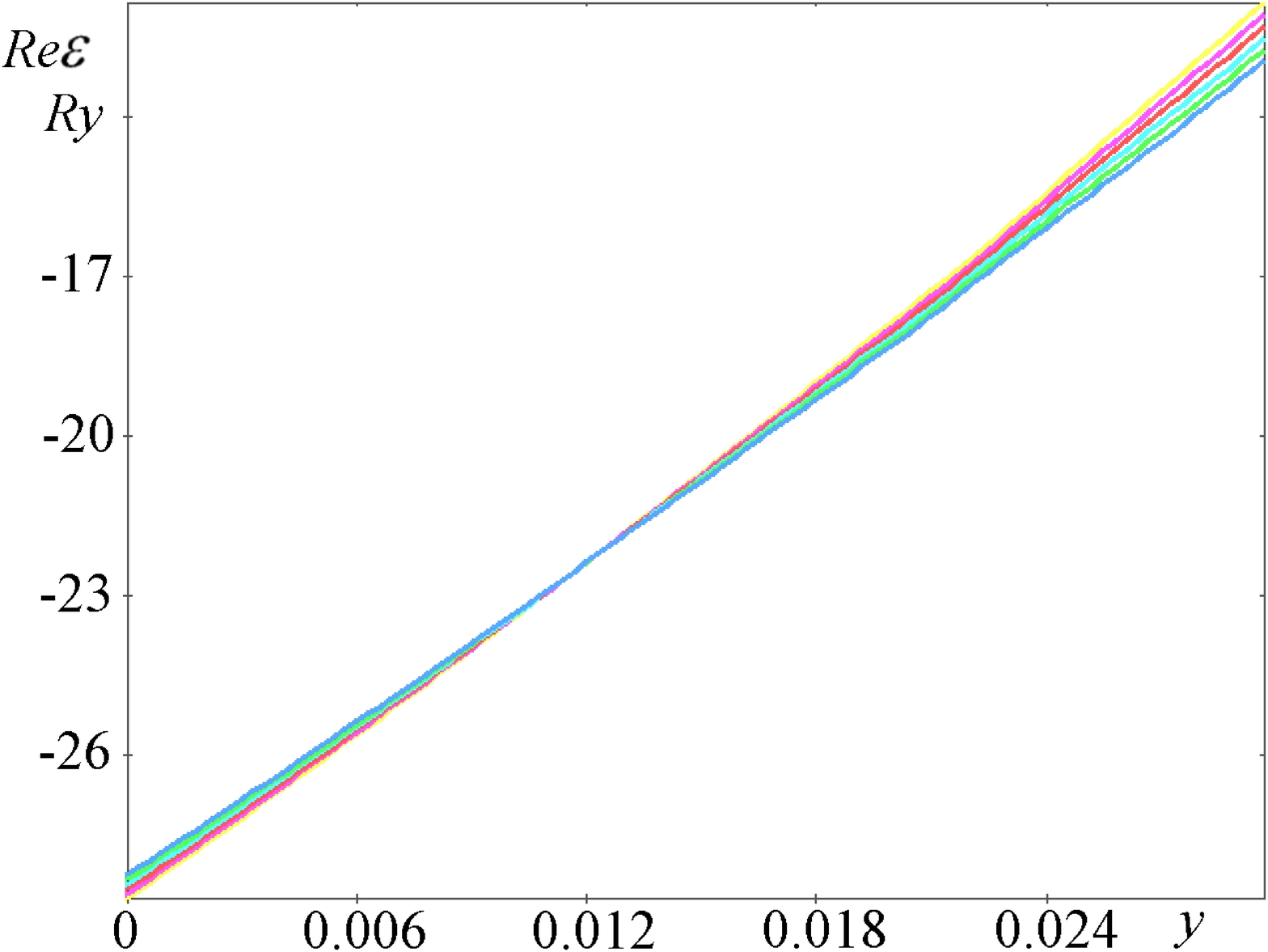}} \caption{
Dependence of the real part of the total energy of a beryllium atom,
$Re\varepsilon$, on the parameter $y$ at $x$ = 0, 0.005, 0.010,
0.015, 0.020, and 0.025. The curve with a smaller slope corresponds
to a larger value of parameter $x$.} \label{figure2}
\end{minipage}
\hfill
\begin{minipage}[!hB]{0.45\linewidth}
\center{\includegraphics[width=1\linewidth]{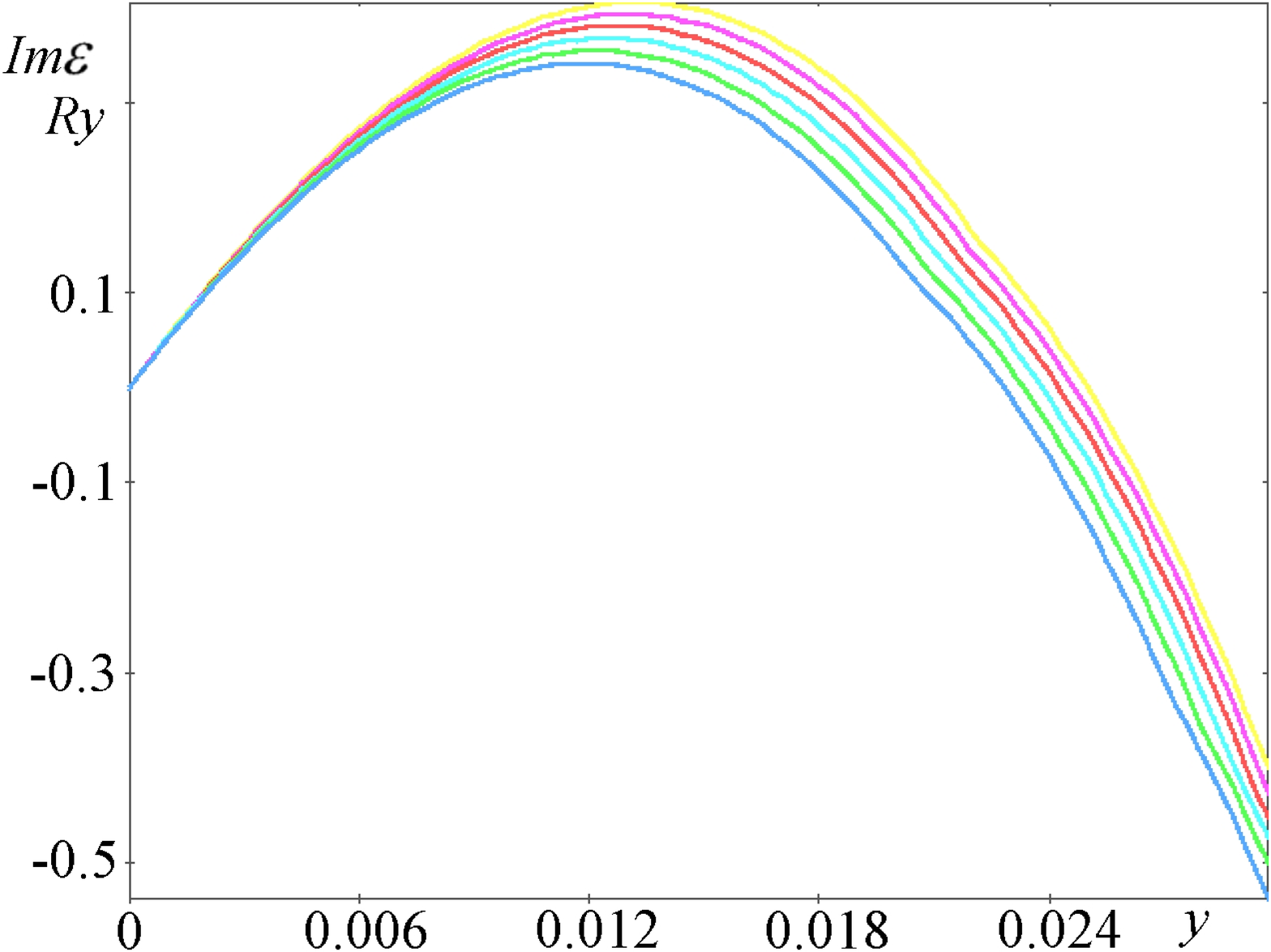}} \caption{
Dependence of the imaginary part of the total energy of a beryllium
atom, $Im\varepsilon$, on the parameter $y$ at $x$ = 0, 0.005,
0.010, 0.015, 0.020, and 0.025. The lower the curve, the larger its
parameter $x$ value.} \label{figure3}
\end{minipage}
\end{figure}

The results of the self-consistent calculations demonstrate that the real part of the total energy of the atom, $Re\varepsilon$, monotonically increases with the parameter $y$ increasing from 0~to~0.03 and shows a weak dependence on $x$ within range from\linebreak 0 to 0.03 (Figure~\ref{figure2}). The imaginary part of the total energy of the atom, $Im\varepsilon$, is shown in Figure~\ref{figure3} as a function of the parameter $y$ for six values of the parameter $x$ = 0, 0.005, 0.010, 0.015, 0.020, and 0.025. Note that the modulus of $Im\varepsilon$ is the probability of the excitation decay per unit time. It is noteworthy that the beryllium atom is stabilized in the region of the decrease of the probability of decay of an excited state at excitation energies higher than 6.7 $Ry$. At least qualitatively, the presence of this region agrees with data from \cite{Delone1995}.

The spectral characteristics of the beryllium atom, such as
single-electron energy levels $E_{n}$, their widths $\Gamma_{n}$,
and probabilities $N_{n}$ of populating of the states, show a weak
dependence on $x$ as well. In essence, the above dependence makes it
possible to trace the variations in the spectral characteristics as
functions of $\Delta\varepsilon=Re(\varepsilon-\varepsilon_{0})$,
which is the excitation energy of an atom with respect to its ground
state $\varepsilon_{0}$, rather than to trace variations in $x$ and
$y$, which are parameters without any physical meaning.

\begin{figure}[!hB]
\center{\includegraphics[width=0.5\linewidth]{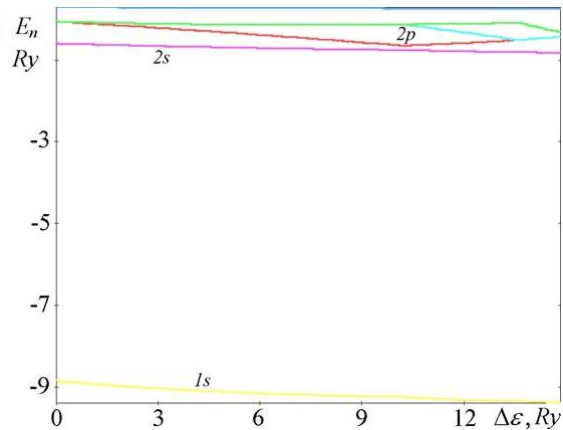}}
\caption{Dependence of the energies of the single-electron states of
a beryllium atom on the excitation energy of an atom.}
\label{figure4}
\end{figure}

Thus, as shown in Figure ~\ref{figure4}, the results of the
self-consistent calculations of energy levels $E_{n}$, which alter
their position as a function of $\varepsilon_{0}$, demonstrate that
the energy of the $1s$ core state decreases with increasing
$\Delta\varepsilon$. On the whole, this agrees with the data from
\cite{Delone1999}, which were experimentally confirmed in
\cite{Bondar}, as well as the behavior of the outer $2s$ state at
small perturbations when the excited states are not yet mixed. Note
that the calculated $1s$, $2s$, $2p$ states still have the
appropriate symmetries, but other numerical values of the energy. So
the excitation changes the positions of the energy levels.

On the whole, the electron can move from $1s$-state to the state of
$2p$-symmetry and to the higher levels, that corresponds to the
X-ray range of excitation energies. By the same the optical
transitions can be attributed to the lower excitation energies, for
example, transitions $2s-2p$. Let us consider the excitation of a
beryllium atom in the optical range of energies. To do this, it is
sufficient to assume that the electron transitions from the $1s$
states are forbidden. In this case, the electron populations of
these states remain unchanged and equal to two (with allowance for
the spin degeneracy). The greatest changes are found in the
occupation numbers of the $2s$ and $2p$ states. The occupation
numbers of the $3s$ states are close to zero, because the
transitions $2s–3s$ are forbidden by the selection rules and the
transitions $2p–3s$ are improbable in view of a small filling of the
$2p$ states by electrons. By increasing of the implicit part of
orbital quantum number $l$ the $1s$ state is raising up in the scale
of energies. This phenomenon is interpreted by us as the atom
collapse, which is intensively discussed in \cite{Delone1995} by the
hydrogen atom.

\begin{figure}[!hB]
\begin{minipage}[!hB]{0.45\linewidth}
\center{\includegraphics[width=1\linewidth]{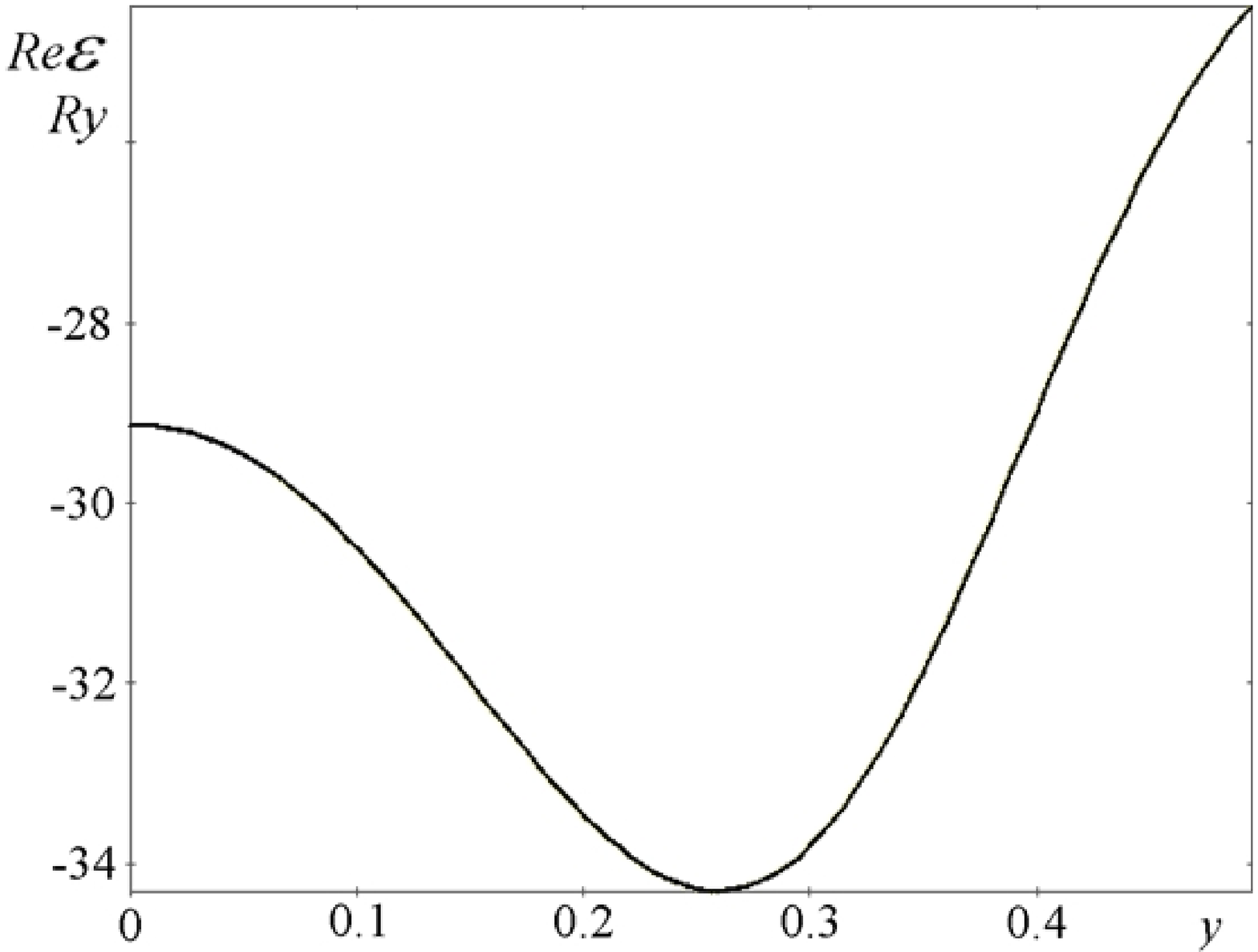} \\ a)}
\end{minipage}
\hfill
\begin{minipage}[!hB]{0.45\linewidth}
\center{\includegraphics[width=1\linewidth]{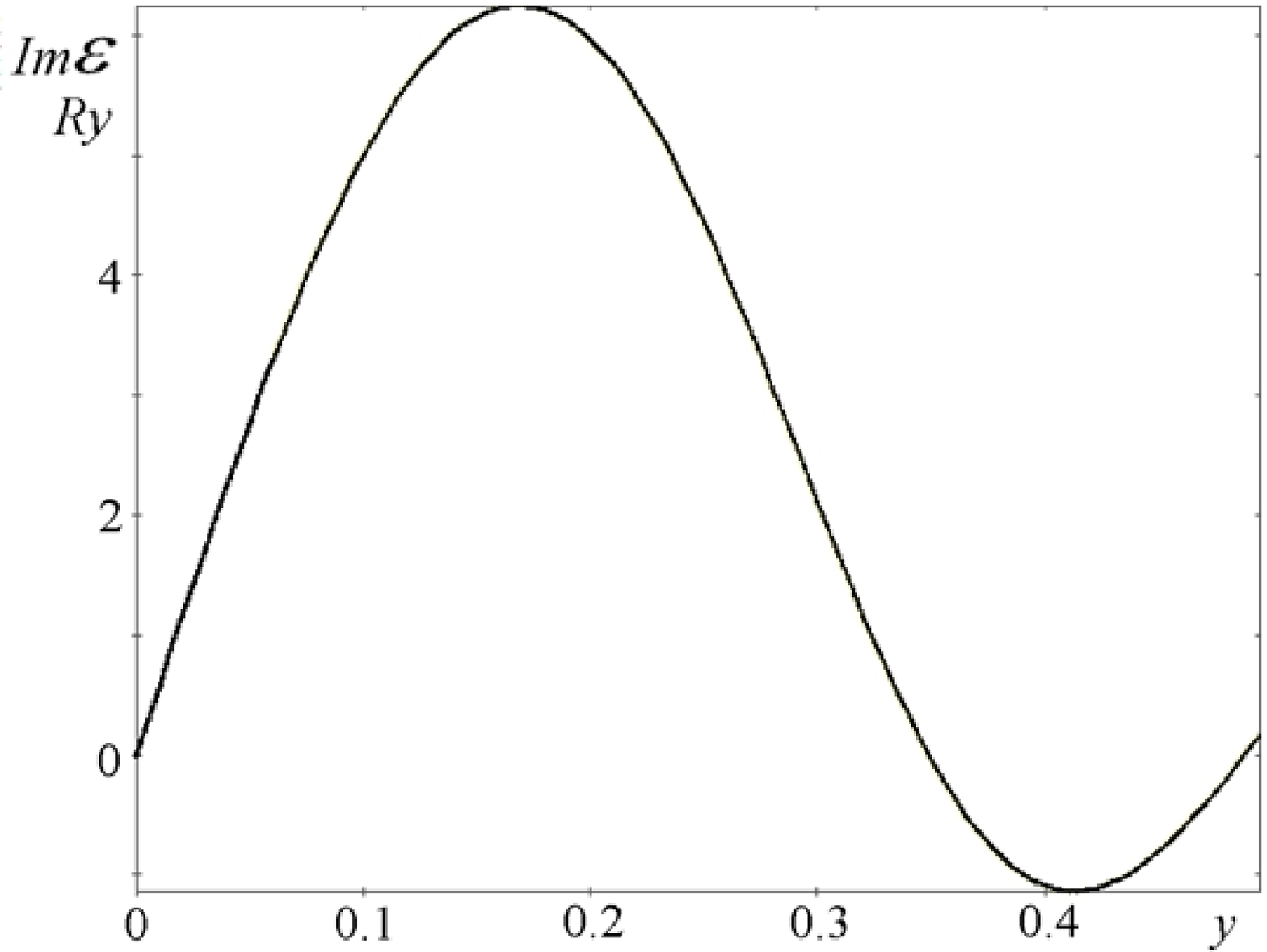} \\ b)}
\end{minipage}
\caption{Dependence of the real part (a) and the imaginary part (b)
of the total energy of a beryllium atom on the parameter $y$ in the
optical range of excitation energies.} \label{figure5}
\end{figure}

Figure~\ref{figure5}a shows the real part $Re\varepsilon$ of the total energy of a beryllium atom calculated as a function of the parameter $y$, which describes excitations in the optical range of energies. These and all other calculations of the excitations in the optical range were carried out for $x=0$ because the imaginary part
of the total energy of the atom is minimal at this value of the parameter $x$ and because the increase of $x$ leads to a corresponding increase in $Re\varepsilon$ at each fixed value of $y$. The dependence shown in Figure~\ref{figure5}a is essentially nonlinear. For $y$ values from 0 to 0.01, the real part $Re\varepsilon$ increases very gradually. Note that, for the sufficiently large $y$ values (from 0.02 to 0.26), the real part of the total energy of the atom decreases considerably and becomes lower than the
energy of the ground state. The lifetimes of an atom in such states are very short because the imaginary part of the total energy, $Im\varepsilon$, which is inversely proportional to the lifetime, is fairly large (see Figure~\ref{figure5}b). The calculated results
illustrated in Figure~\ref{figure5}b show that the imaginary part vanishes, $Im\varepsilon=0$, at certain $y$ values, in particular, at $y=0.35$ and 0.49. This indicates that there exist at least two long-lived states of a highly excited atom. One of these excited states, namely, that with $y=0.35$, is stable because its total energy is lower than the energy of the ground state at $y=0$. It is
very unlikely, however, that, in the optical range, an atom can be excited to such a high state (with $y=0.35$). With the strengths of the fields used in practice, it is only possible to change the parameter $y$ within a few thousandths or, at least, a few hundredths. Nevertheless, it is possible to make an excited state stable, e.g., by placing the atom in the field of another atom \cite{Popov2006}. To test this hypothesis, we calculated the energy structure of a few atoms, the results of which will be presented in the next section.

\section{Small beryllium clusters in intense fields}
\label{sec3}

\subsection{Computational technique}

Let us generalize the above proposed approach to the description of
excited states of polyatomic systems, such as clusters. The total
energy of all electrons in the Hartree-Fock approximation can be
represented in the form of the sum
\begin{equation}
\label{6}
\varepsilon_{T}=\varepsilon_{K}+\varepsilon_{N}+\varepsilon_{E}+
\varepsilon_{C}+\varepsilon_{X}+\varepsilon_{L},
\end{equation}
which contains the kinetic energy of electrons
\begin{equation}
\label{7} \varepsilon_{K}=\sum \limits_{i}^{occ}\int
\psi_{i}^{*}(\textbf{r})\left(-\frac{\hbar^{2}}{2m}
\Delta\right)\psi_{i}(\textbf{r})d^{3}r,
\end{equation}
the potential energy of the electron–nucleus Coulomb interaction
\begin{equation}
\label{8} \varepsilon_{N}=-\sum
\limits_{i}^{occ}\int\psi_{i}^{*}(\textbf{r})\sum
\limits_{I}\frac{Z_{I}e^{2}}{|\textbf{r}+\textbf{a}_{I}|}
\psi_{i}(\textbf{r})d^{3}r,
\end{equation}
the potential energy of the electron-electron Coulomb interaction
\begin{equation}
\label{9} \varepsilon_{E}=\sum
\limits_{i}^{occ}\int\psi_{i}^{*}(\textbf{r})\psi_{i}(\textbf{r})
\int\frac{\rho(\textbf{r}')e^{2}}
{|\textbf{r}-\textbf{r}'|}d^{3}r'd^{3}r,
\end{equation}
the potential energy of the nuclei Coulomb interaction
\begin{equation}
\label{10} \varepsilon_{C}=\frac{1}{2}\sum \limits_{I,J\neq
I}\frac{Z_{I}Z_{J}e^{2}}{|\textbf{a}_{I}-\textbf{a}_{J}|},
\end{equation}
and the energy of the electron exchange interaction
\begin{equation}
\label{11} \varepsilon_{X}=e^{2}\sum \limits_{i}^{occ}\sum
\limits_{i'}^{occ}\int\int\psi_{i}^{*}(\textbf{r})\psi_{i}(\textbf{r}')
\psi_{i'}^{*}(\textbf{r})\psi_{i'}(\textbf{r})\frac{d^{3}r'd^{3}r}
{|\textbf{r}-\textbf{r}'|}.
\end{equation}
Here, the electron density of a cluster
\begin{equation}
\label{12} \rho(\textbf{r})=\sum \limits_{i}^{occ}
|\psi_{i}(\textbf{r})|^{2}
\end{equation}
is the sum over all occupied states with numbers $i$. The last term in the relation (\ref{6}) for the total energy can be written, due to our idea, in the form
\begin{equation}
\label{13} \varepsilon_{L}=\sum \limits_{i}^{occ}\int
\psi_{i}^{*}(\textbf{r})\left(\frac{\hbar^{2}}{2m}\frac{u+iv}{r^{2}}
\right)\psi_{i}(\textbf{r})d^{3}r,
\end{equation}

The wave function can be expanded over the complete set of functions
${\varphi_{n}(\textbf{r})}$:
\begin{equation}
\label{14} \psi(\textbf{r})=\sum
\limits_{j,n}C_{ij}\left(\textbf{S}^{-1/2}\right)_{jn}\varphi_{n}(\textbf{r}).
\end{equation}
Here, the overlap integral matrix $\textbf{S}$ differs from a unit
matrix if the functions $\varphi_{n}(\textbf{r})$ are not
orthonormalized. The functions $\varphi_{n}(\textbf{r})$ are
represented by Gaussian functions:
\begin{equation}
\label{15}
\varphi_{n}(\textbf{r})=\left(\frac{2\alpha_{n}}{\pi}\right)^{3/4}
exp\left(-\alpha_{n}|\textbf{r}-\textbf{a}_{n}|^{2}\right),
\end{equation}
where $\textbf{a}_{n}$ is the vector of the atomic position with respect to the chosen coordinate system. It is important that the functions with symmetries $p, d, f, ...$ can be easily obtained from functions (\ref{15}) with the symmetry $s$ by differentiation with respect to the components of the vector $\textbf{a}_{n}$. This idea underlies the analytical calculations of the matrix elements. Taking into account this circumstance, the main
attention can be focused on the derivation of the matrix elements with the symmetry $s$.

The necessary condition for the existence of a minimum of the total-energy functional (\ref{6}) is zero of its first variation. In the Roothaan scheme, the solution of this problem is reduced to the determination of the variation coefficients $C_{ij}$ entering into Eq.~(\ref{6}). In order to solve the system of equations obtained in the Roothaan scheme, it is necessary to
calculate the matrix elements in the explicit form. For states with
the symmetry $s$, the matrix element of the overlap integrals is
represented in the following form:
\begin{equation}
\label{16}
S_{12}\equiv\int\varphi_{1}^{*}(\textbf{r})\varphi_{2}(\textbf{r})d^{3}r=
\left(\frac{2\sqrt{\alpha_{1}\alpha_{2}}}{\alpha_{1}+\alpha_{2}}\right)^{3/2}
exp\left(-\frac{\alpha_{1}\alpha_{2}}{\alpha_{1}+\alpha_{2}}
|\textbf{a}_{1}-\textbf{a}_{2}|^{2}\right).
\end{equation}

The matrix element of the kinetic energy operator is written in the
form
\begin{eqnarray}
\label{17} K_{12}\equiv\int
\varphi_{1}^{*}(\textbf{r})\left(-\frac{\hbar^{2}}{2m}\Delta\right)
\varphi_{2}(\textbf{r})d^{3}r=\nonumber\\
=\frac{\hbar^{2}}{2m}S_{12}\frac{\alpha_{1}\alpha_{2}}{\alpha_{1}+\alpha_{2}}
\left(6-\frac{4\alpha_{1}\alpha_{2}}{\alpha_{1}+\alpha_{2}}
|\textbf{a}_{1}-\textbf{a}_{2}|^{2}\right).
\end{eqnarray}

The matrix element of the electron–nucleus interaction has the form
\begin{eqnarray}
\label{18} N_{12}\equiv-e^{2}\sum\limits_{i}Z_{i}\int
\varphi_{1}^{*}(\textbf{r})\varphi_{2}(\textbf{r})
\frac{d^{3}r}{|\textbf{r}-\textbf{a}_{i}|}=\nonumber\\
=-e^{2}S_{12}\sum\limits_{i}Z_{i}\frac{erf(R_{i}
\sqrt{\alpha_{1}+\alpha_{2}})}{R_{i}}.
\end{eqnarray}
In Eq.~(\ref{18}), $erf(x)$ is the error function and

\begin{equation}
\label{19}
\textbf{R}_{i}\equiv\frac{\alpha_{1}\textbf{a}_{1}+\alpha_{2}\textbf{a}_{2}}
{\alpha_{1}+\alpha_{2}}-\textbf{a}_{i}.
\end{equation}

The matrix element of the electron–electron interaction is given by the expression
\begin{eqnarray}
\label{20} E_{1234}\equiv e^{2}\int
\varphi_{1}^{*}(\textbf{r})\varphi_{2}(\textbf{r})
\varphi_{3}^{*}(\textbf{r}')\varphi_{4}(\textbf{r}')=\nonumber\\
=e^{2}S_{12}S_{34}\frac{erf(\textbf{R}\sqrt{\beta})}{\textbf{R}},
\end{eqnarray}
where
\begin{equation}
\label{21}
\textbf{R}\equiv\frac{\alpha_{1}\textbf{a}_{1}+\alpha_{2}\textbf{a}_{2}}
{\alpha_{1}+\alpha_{2}}-
\frac{\alpha_{3}\textbf{a}_{3}+\alpha_{4}\textbf{a}_{4}}
{\alpha_{3}+\alpha_{4}},
\end{equation}
\begin{equation}
\label{22}
\beta\equiv\frac{(\alpha_{1}+\alpha_{2})(\alpha_{3}+\alpha_{4})}
{\alpha_{1}+\alpha_{2}+\alpha_{3}+\alpha_{4}}.
\end{equation}

The matrix element of the square of the angular momentum is defined
as
\begin{eqnarray}
\label{23} M_{12}\equiv\frac{\hbar^{2}}{2m} \int
\varphi_{1}^{*}(\textbf{r})\varphi_{2}(\textbf{r})\frac{d^{3}r}{r^{2}}=\nonumber\\
=\frac{\hbar^{2}}{2m}\frac{\pi^{2}}{\alpha}
\left(\frac{2\sqrt{\alpha_{1}\alpha_{2}}}{\pi}\right)^{3/2}
exp(-\alpha_{1}\textbf{a}_{1}^{2}-\alpha_{2}\textbf{a}_{2}^{2})
\frac{erf(i\sqrt{\alpha}\textbf{a})}{i\textbf{a}},
\end{eqnarray}
where
\begin{equation}
\label{24}
\textbf{a}\equiv\frac{\alpha_{1}\textbf{a}_{1}+\alpha_{2}\textbf{a}_{2}}
{\alpha}, \qquad \alpha=\alpha_{1}+\alpha_{2}.
\end{equation}

It should be noted that the matrix elements (\ref{23}) can be used
when the many-electron problem has a central symmetry at
$\textbf{a}_{1}=\textbf{a}_{2}=0$. In the case of a multicenter
problem, for example, for a cluster, the matrix elements of the
square of the angular momentum can be calculated from the formula:
\begin{equation}
\label{25}
L_{12}\equiv\sum\limits_{ijn}(\textbf{S}^{-1/2})_{1n}
\sum\limits_{k}B_{nk}^{+}\Lambda_{k}'
B_{ki}(\textbf{S}^{-1/2})_{ij}M_{j2}.
\end{equation}
Here, the quantities $B_{ki}$ have the meaning of the components of the eigenvectors of the spherical Laplace operator. The matrix
elements calculated for the spherical Laplace operator with the use
of functions (\ref{15}) with the symmetry $s$ can be represented in terms of a function of the projections of the vectors $\textbf{a}_{1}$ and
$\textbf{a}_{2}$ onto the axes of the Cartesian coordinate system in
the following form:
\begin{eqnarray}
\label{26} \Lambda_{12}\equiv\int\varphi_{1}^{*}(\textbf{r})\Delta
\varphi_{2}(\textbf{r}) d^{3}r=\nonumber\\
=-4S_{12}\left(\frac{\alpha_{1}\alpha_{2}}{\alpha_{1}+\alpha_{2}}
\right)
(\alpha_{1x}\alpha_{2x}+\alpha_{1y}\alpha_{2y}+\alpha_{1z}\alpha_{2z})+\nonumber\\
+4S_{12}\left(\frac{\alpha_{1}\alpha_{2}}{\alpha_{1}+\alpha_{2}}\right)^{2}
(|\alpha_{1x}\alpha_{2y}-\alpha_{2x}\alpha_{1y}|^{2}+\\
+|\alpha_{1x}\alpha_{2z}-\alpha_{2x}\alpha_{1z}|^{2}+
|\alpha_{1y}\alpha_{2z}-\alpha_{2y}\alpha_{1z}|^{2}).\nonumber\end{eqnarray}

The change $\Lambda_{k}'$ in the $k$-th eigenvalue of the spherical
Laplace operator in Eq.~(\ref{25}) can be expressed in terms of a function of the quantity $\Lambda_{k}$ (the $k$-th eigenvalue of this operator):
\begin{equation}
\label{27} \Lambda_{k}'=(l_{k}+x+iy)(l_{k}+x+iy+1)-\Lambda_{k},
\qquad l_{k}=\sqrt{0.25+\Lambda_{k}}-0.5.
\end{equation}

The expression for the total energy of the cluster can be
represented from the matrix elements determined above:
\begin{equation}
\label{28}
\varepsilon_{T}=\sum\limits_{n,m}(K_{nm}+N_{nm}+L_{nm}+F_{nm})U_{nm}+
\varepsilon_{C}.
\end{equation}
Here, the matrix elements
\begin{equation}
\label{29} U_{nm}=\sum\limits_{k}^{occ}C_{ki}^{+}C_{kj}
\sum\limits_{i,j}(\textbf{S}^{-1/2})_{1n}(\textbf{S}^{-1/2})_{jm},
\end{equation}
\begin{equation}
\label{30} F_{nm}=K_{nm}+N_{nm}+L_{nm}+
\sum\limits_{i,j}(2E_{nmij}-E_{nimj}U_{ij})
\end{equation}
are necessary for the self-consistent search for the solutions to
the algebraic eigenvalue problem in the matrix form:
\begin{equation}
\label{31} \textbf{FC}=\textbf{ESC},
\end{equation}
where $\textbf{E}$ is the diagonal matrix of eigenvalues,
$\textbf{C}$ is the matrix of eigenvectors, $\textbf{F}$ is the
matrix with elements (\ref{30}), and $\textbf{S}$ is the overlap
integral matrix with elements (\ref{16}).

\subsection{Some practical aspects of calculations}

It was noted that the functions with symmetries $p, d, f, ...$ can be easily obtained using functions (\ref{15}) with the symmetry $s$ by differentiation with respect to the components of the vector $\textbf{a}_{n}$.

Calculation of $\textbf{S}^{-1/2}$ is convenient to perform in a form (\ref{32}), according to the Cayley-Hamilton theorem:
\begin{equation}
\label{32}
(\textbf{S}^{-1/2})_{ij}=\sum\limits_{k}\lambda_{k}^{-1/2}c_{ki}c_{kj}.
\end{equation}

Here $\lambda_{k}$ is the $k$-th eigenvalue and the corresponding
eigenvector ${c_{k1}, c_{k2}, ...}$ of the positive definite overlap
integral matrix. For search of the eigenvalues and eigenvectors we
used a computational Householder scheme \cite{Householder}. The Jacobi method for complex matrices \cite{Anderson} was used in the solution of the spectral problem (\ref{31}) at every stage of the iterative search for a self-consistent solution.

The value $U_{nm}=0$ was chosen as a zero approximation of the iterative search for a self-consistent solutions. The calculated matrix $\textbf{F}$ was used for the solution of the generalized eigenvalue problem (\ref{31}). Found in the first iteration, the eigenvectors $\textbf{C}$ were used to calculate $U_{nm}$ from (\ref{29}) to form the matrix $\textbf{F}$ for solving the generalized eigenvalue problem (\ref{31}) at the next stage of the self-consistency procedure. The iterative procedure of searching of the self-consistent solutions was continued until the eigenvalues of the diagonal matrix $\textbf{E}$, found at the previous step, coincided with the desired accuracy of $10^{-5}$ with the eigenvalues of the diagonal matrix, found at the subsequent stage. Note that this procedure does not always converge, because the values ${U_{nm}}^{(i-1)}$, found at the $(i-1)$-th stage may differ from the values of $U_{nm}^{(i)}$ found at the $(i)$-th stage. There are
many ways to avoid these divergences. We used the simplest and the most efficient of them, when instead of $U_{nm}^{(i)}$ the value, calculated using the formula $\eta U_{nm}^{(i)}+(1-\eta){U_{nm}}^{(i-1)}$, was used. Here, the damping factor $\eta$ is chosen for the practical reasons, $0<\eta \leq1$.

\subsection{Results of calculations}

Problem (\ref{31}) was numerically solved in the basis set of
functions of the Gaussian type with the use of nine functions in the
expansion in $l=0$, six functions in the expansion in $l=1$, and
three functions in the expansion in $l=2$. The estimates
demonstrated that the basis set of this size is quite suitable for
solving Eqs. \ref{31} using the Roothaan method for the beryllium atom
at the parameters $|x|<0.5$ and $|y|<0.5$, which determine the change in
the orbital angular momentum in relationships (\ref{27}). It should
be noted that the size increase of this basis set does not change
the required accuracy of all values given below for the discussion.
The algorithm for solving the aforementioned equations is well known
\cite{Baranovskioe}.

\begin{figure}[!hB]
\begin{minipage}[!hB]{0.45\linewidth}
\center{\includegraphics[width=0.9\linewidth]{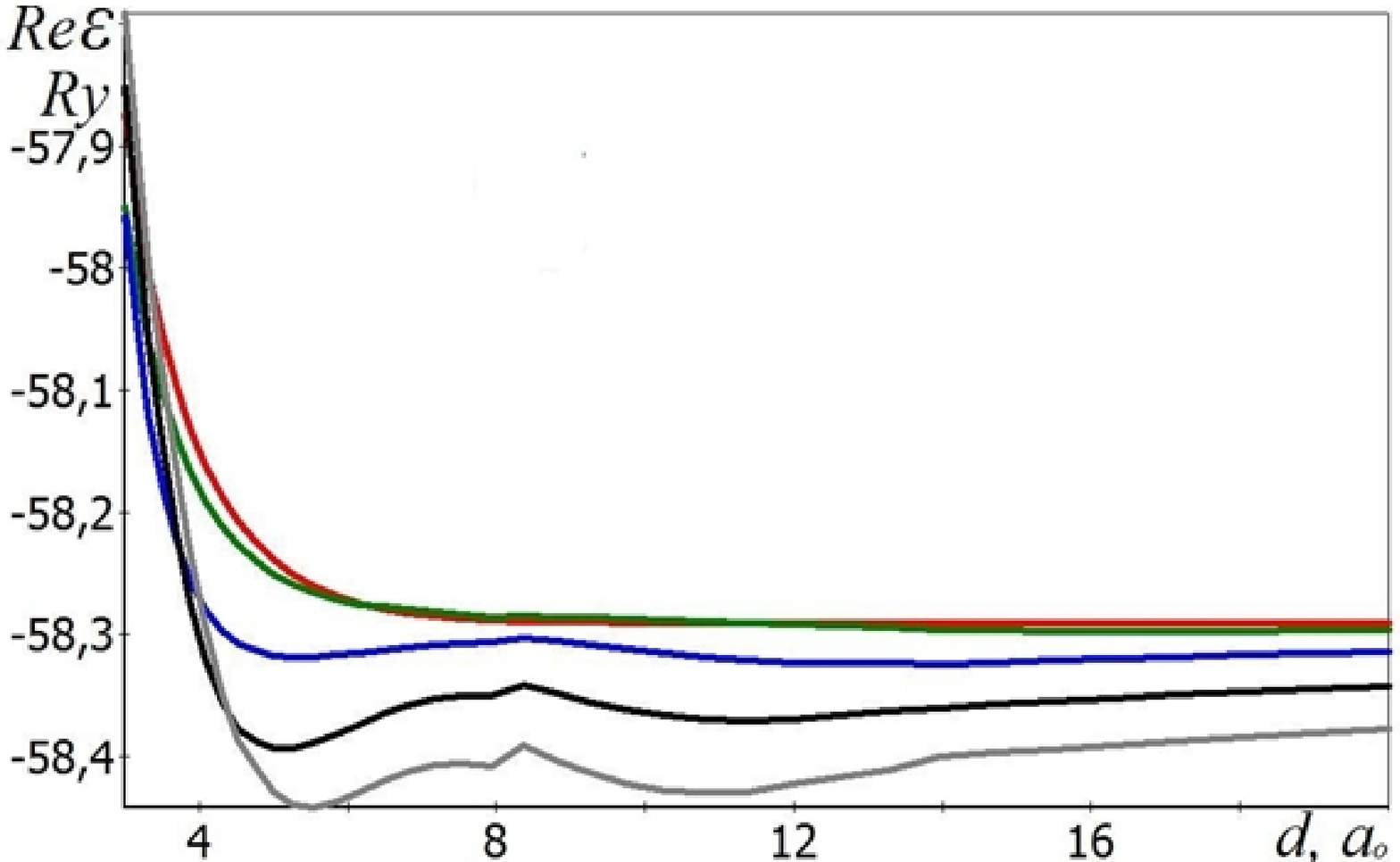}}
\caption{ Dependence of the real part of the total electron energy
for two beryllium atoms on the interatomic distance $d$ at the
parameters $x=0$ and $y=0, 0.0001, 0.0002, 0.0003,$ and 0.0004 in
the optical range of excitation energies. The lower the location of
the curve, the larger the parameter $y$ corresponding to this
curve.} \label{figure6}
\end{minipage}
\hfill
\begin{minipage}[!hB]{0.45\linewidth}
\center{\includegraphics[width=0.9\linewidth]{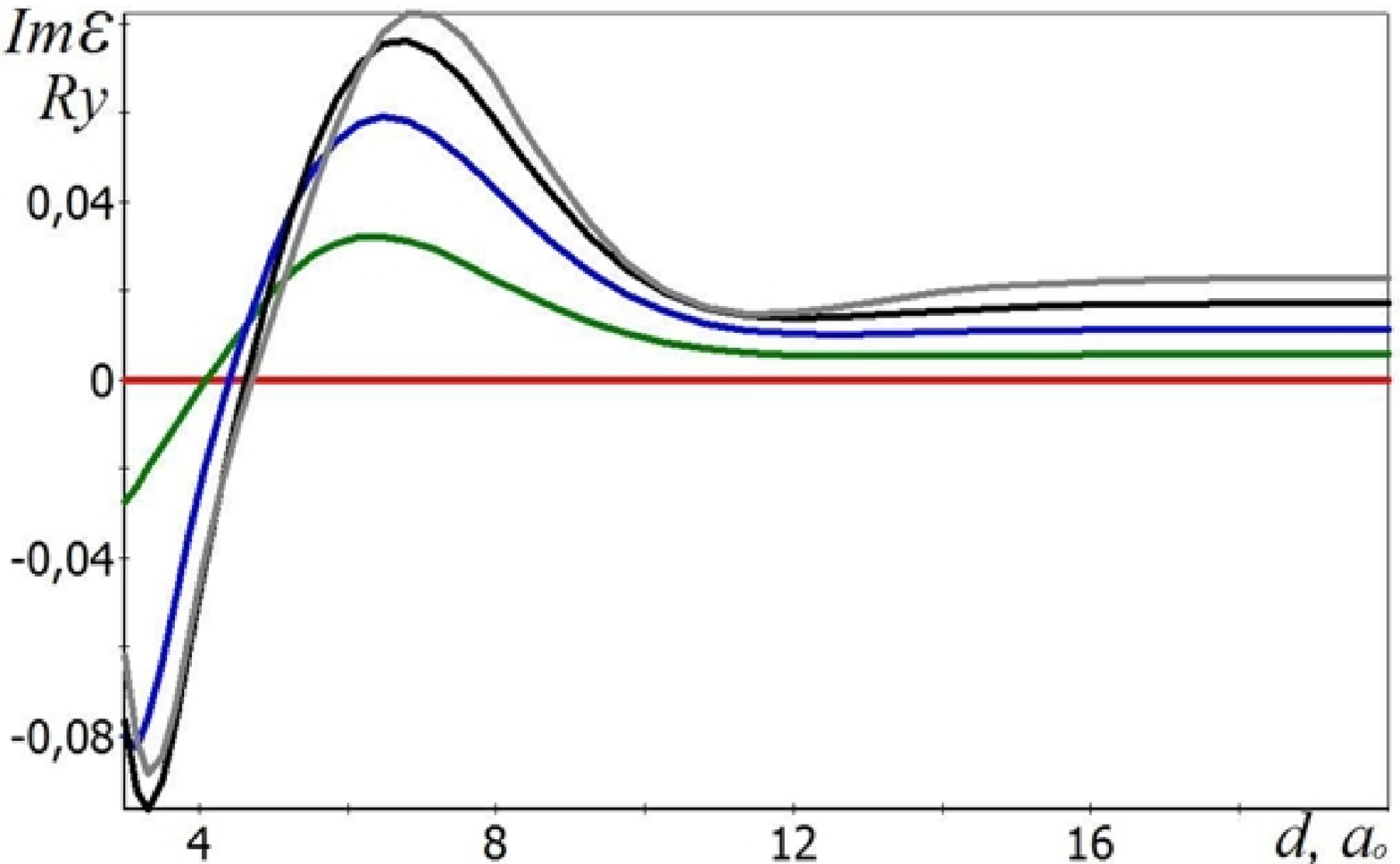}}
\caption{ Dependence of the imaginary part of the total electron
energy for two beryllium atoms on the interatomic distance $d$ at
the parameters $x=0$ and $y=0, 0.0001, 0.0002, 0.0003,$ and 0.0004
in the optical range of excitation energies. Curves with larger
deviations from the straight line $Im\varepsilon=0$ correspond to
larger values of $y$.} \label{figure7}
\end{minipage}
\end{figure}

We have calculated the total energy of two beryllium atoms as a
function of the interatomic distance at small parameters $y=0,
0.0001, 0.0002, 0.0003,$ and 0.0004 in the optical range of
excitation energies. The results of calculations (presented in
Figure~\ref{figure6}) indicate that the energy of the ground state
of the system formed by two atoms (the curve with $y=0$ in
Figure~\ref{figure6} at all interatomic distances in the range from
three to twenty Bohr radii is higher than the energy of the ground
state of two isolated atoms. This implies that the $Be_{2}$ system
in the ground state is unstable. At any arbitrarily small
excitations with $y>0$, the total energy of two beryllium atoms
decreases. The larger the parameter $y$, the larger this decrease.
It is worth noting that there are two minima in the dependence of
the real part $Re\varepsilon$ of the total energy at interatomic
distances on the order of five and eleven Bohr radii. A decrease in
the imaginary part $Im\varepsilon$ of the total energy of the system
formed by two beryllium atoms with a decrease in the interatomic
distance (Figure~\ref{figure7}) suggests a stabilization of the
system. An insignificant but clearly pronounced minimum in the
dependence of the quantity $Im\varepsilon$ corresponds to the same
interatomic range in the vicinity of eleven bohr. The range with
$Im\varepsilon=0$ at interatomic distances slightly larger than four
Bohr radii is most interesting. At these distances, experimentally
observed states with an infinitely long lifetimes can be formed in
the $Be_{2}$ system. It should be noted that S. Sharma et al.
\cite{Sharma} conducted a precision calculations of beryllium dimer
in the ground and excited states, which showed the formation of
$Be_{2}$ at a distance of 4.15 bohr, which is much smaller than the
experimental values at 4.62 bohr \cite{Merritt}. This discrepancy
with the experiment was explained by inaccuracies in the conduct of
the experiment. Besides, a metastable state of beryllium dimer with
the distance between the atoms of about 4.2 bohr was mentioned in
\cite{Matxain}. In general, the described phenomenon in beryllium
can be referred to as photocondensation \cite{Bezuglyi}.

\begin{figure}[!hB]
\begin{minipage}[!hB]{0.45\linewidth}
\center{\includegraphics[width=0.9\linewidth]{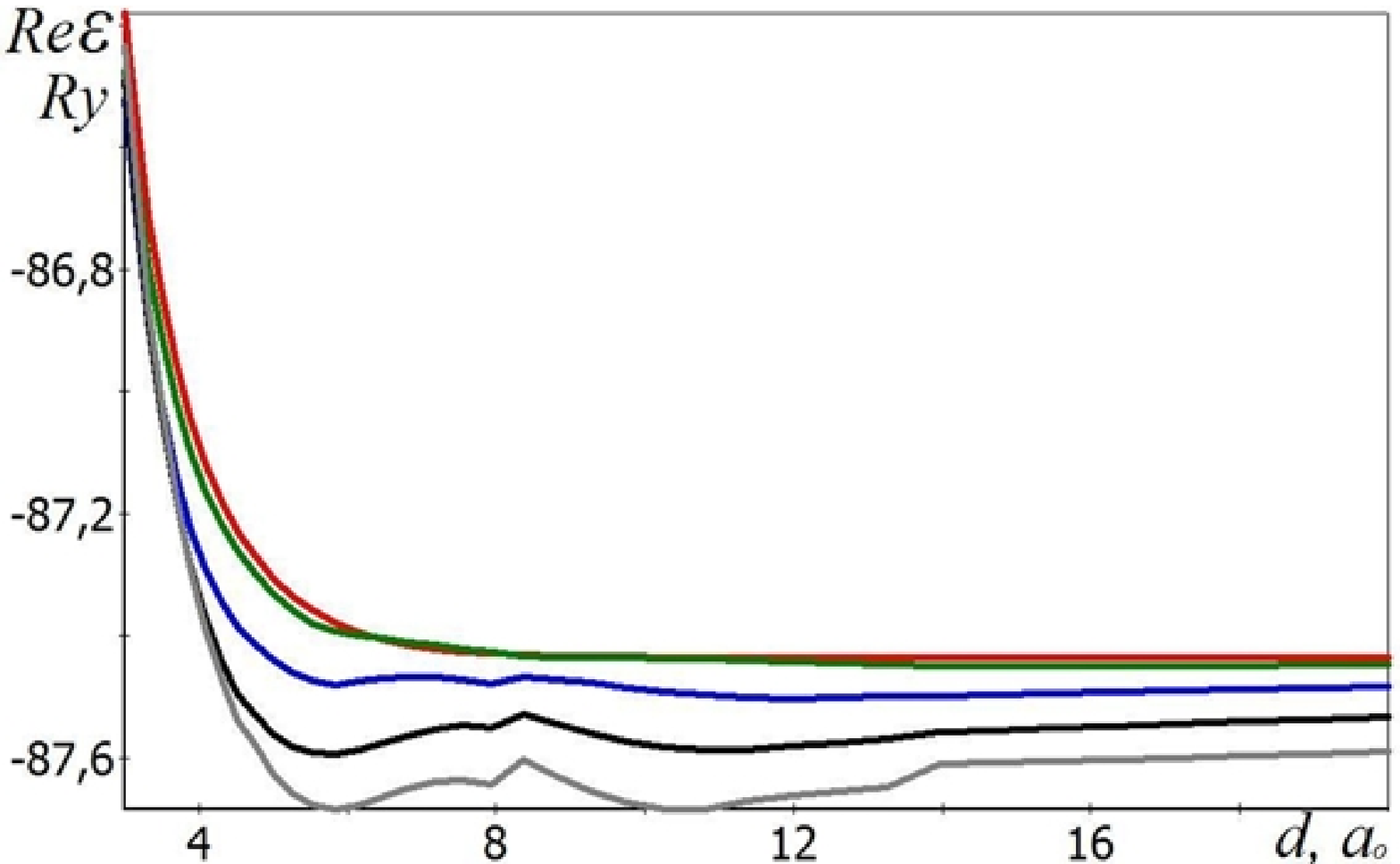}}
\caption{ The same as in Fig. 6 but for three beryllium atoms.}
\label{figure8}
\end{minipage}
\hfill
\begin{minipage}[!hB]{0.45\linewidth}
\center{\includegraphics[width=0.9\linewidth]{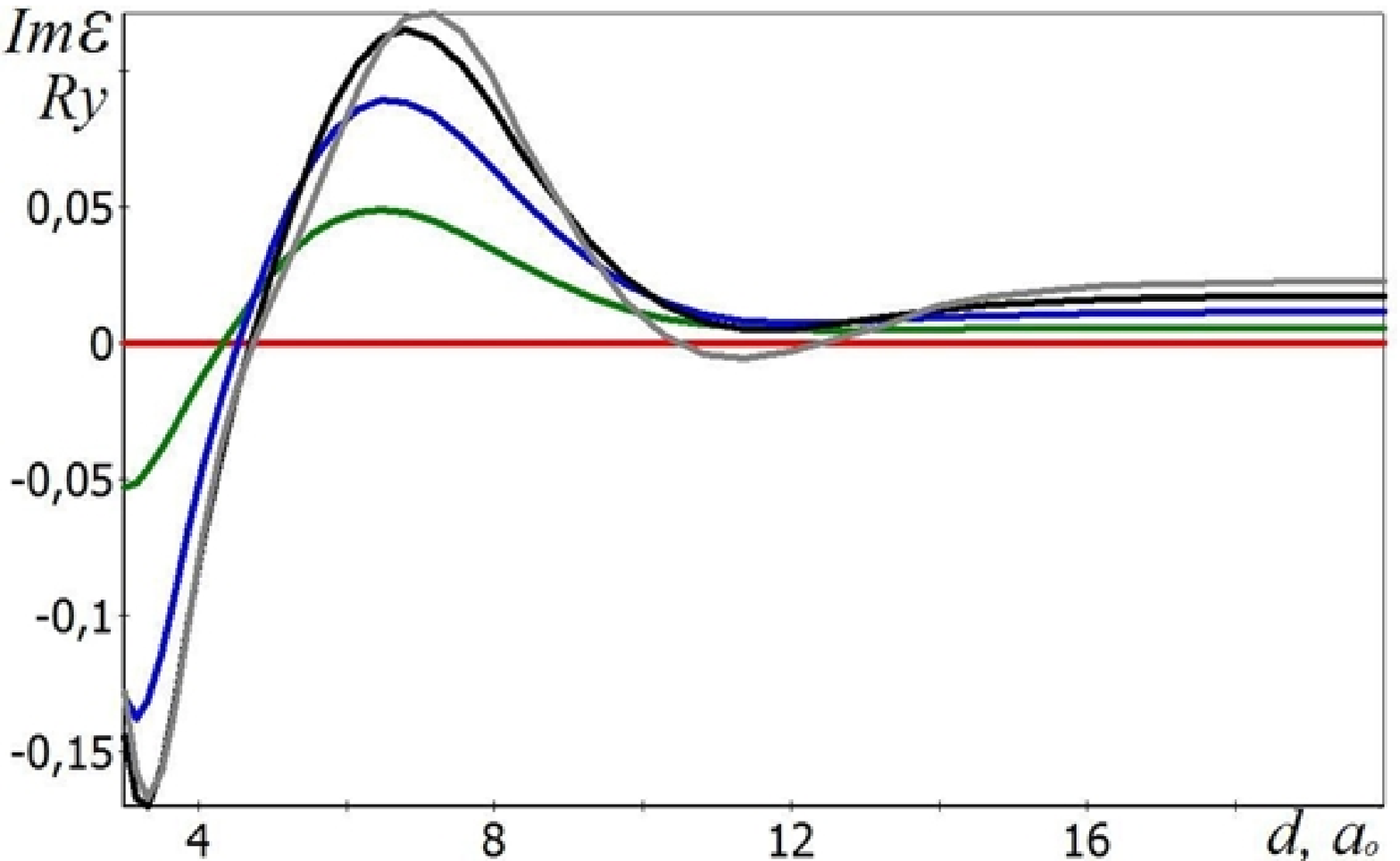}}
\caption{ The same as in Fig. 7 but for three beryllium atoms.}
\label{figure9}
\end{minipage}
\end{figure}

\begin{figure}[!hB]
\begin{minipage}[!hB]{0.45\linewidth}
\center{\includegraphics[width=0.9\linewidth]{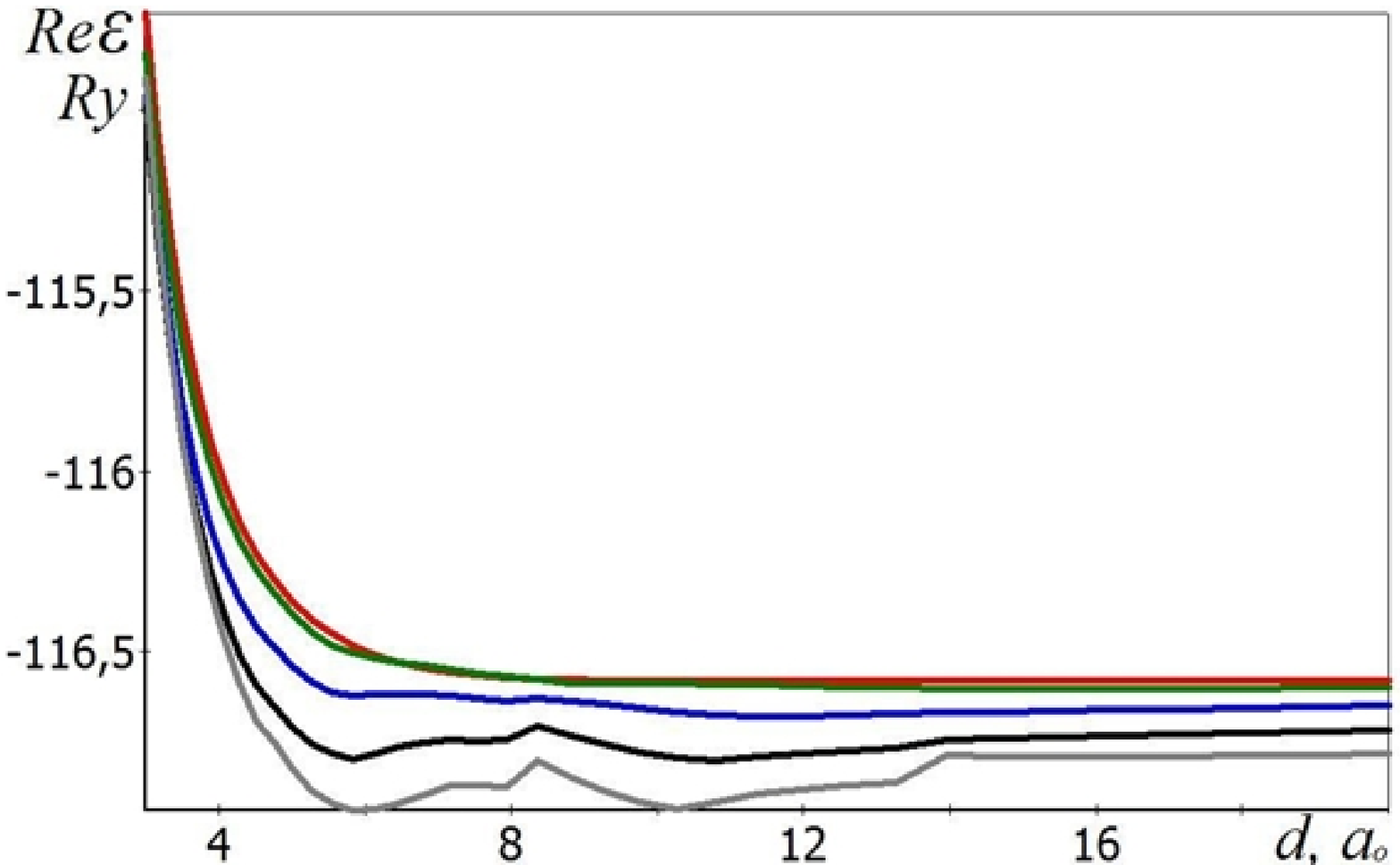}}
\caption{ The same as in Fig. 6 but for four beryllium atoms.}
\label{figure10}
\end{minipage}
\hfill
\begin{minipage}[!hB]{0.45\linewidth}
\center{\includegraphics[width=0.9\linewidth]{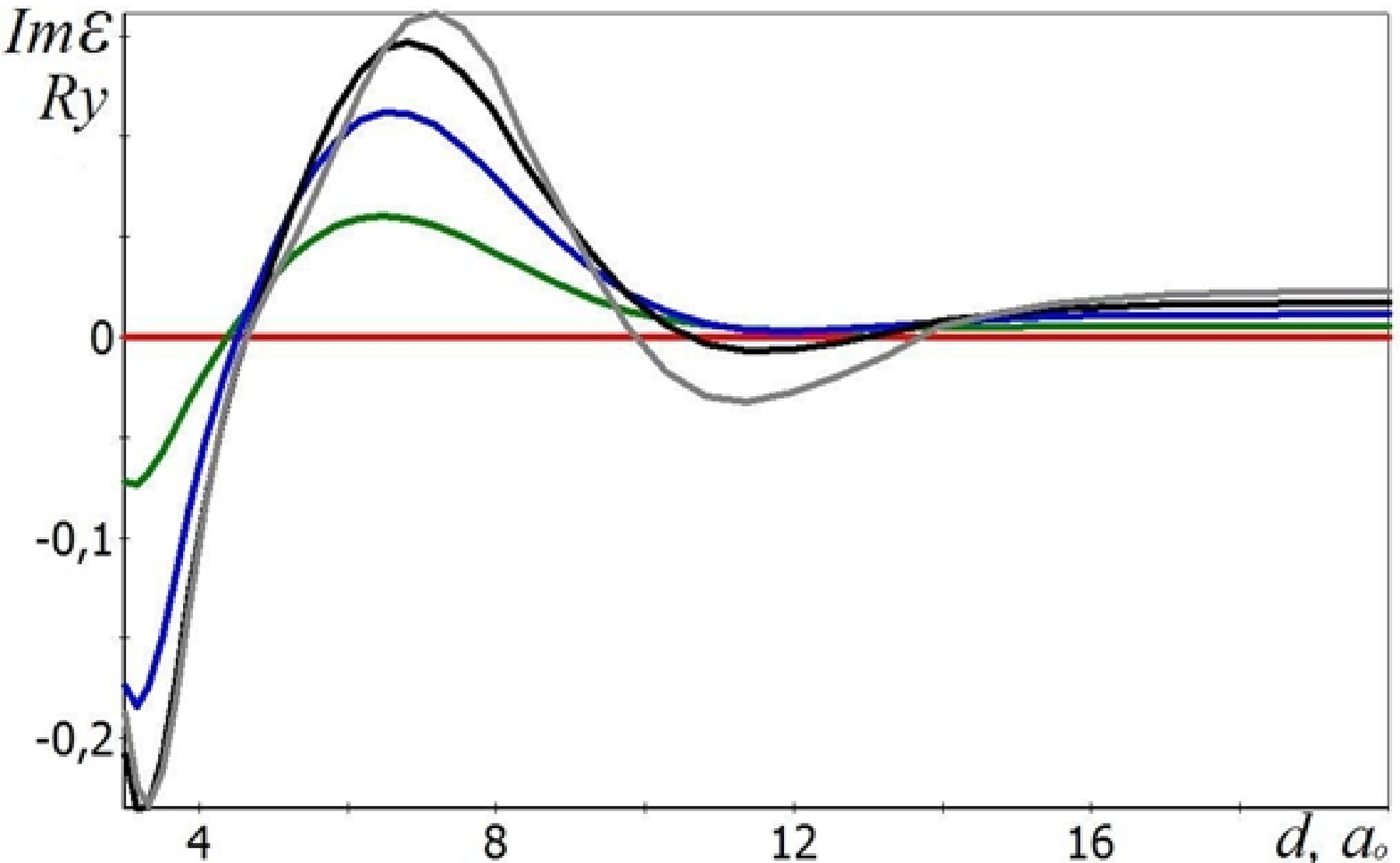}}
\caption{ The same as in Fig. 7 but for four beryllium atoms.}
\label{figure11}
\end{minipage}
\end{figure}

The obtained data permit us to assume that, if the beryllium atom will be placed in a field produced by two, three, or a larger number of atoms, then a cluster can be form more stable in the energy with a
longer lifetime as compared to the diatomic cluster at interatomic
distances in the vicinity of eleven bohr. In order to verify this
assumption, we calculated the total energy of three beryllium atoms
as a function of the interatomic distance at the same parameters
$x=0$ and $y=0, 0.0001, 0.0002, 0.0003,$ and 0.0004 in the optical
range of excitation energies for the optimum geometry of the atoms
located at the vertices of an equilateral triangle. The results of
calculations (presented in Figure~\ref{figure8}) indicate that the
behavior of the real part of the total energy is qualitatively
similar to the corresponding behavior for the system of two atoms.
However, the behavior of the imaginary part of the total electron
energy (Figure~\ref{figure9}) implies not only a stabilization of
three beryllium atoms but also the presence of long-lived
excitations in the $Be_{3}$ system at interatomic distances on the
order of eleven bohr. As for the $Be_{2}$ system, infinitely
long-lived states can be formed at distances slightly larger than
four bohr in the $Be_{3}$ system.

In order to verify the revealed specific features, we calculated the
total energy of four beryllium atoms as a function of the
interatomic distance at the same parameters $x=0$ and $y=0, 0.0001,
0.0002, 0.0003,$ and 0.0004 in the same optical range of excitation
energies. The results of calculations presented in
Figure~\ref{figure10} and Figure~\ref{figure11} only quantitatively
differ from those presented in Figure~\ref{figure8} and
Figure~\ref{figure9} and confirm our assumptions. The formation of
long-lived states at interatomic distances in the vicinity of eleven
bohr implies that there is a strong electron correlation in the
beryllium clusters under conditions of external excitations. We
believe that the revealed behavior of strongly correlated Fermi
systems is also inherent in other clusters \cite{Shaginyan}.

Note that the presence of the second minimum in small beryllium
clusters was mentioned in \cite{Pacchioni}. $Be_{7}$ cluster with
the distance between atoms near 4.2 bohr and also near 9 bohr was
found in this paper. The data for the ground state is in qualitative
agreement with our data. Moreover calculations for $Be_{4}$,
presented in Figure~\ref{figure11}, show that the minimum of the
total energy shifts towards the closer values of distances between
atoms with the increase of external influence. This, in turn, is
consistent with the work \cite{Mel'nikov}, where was shown that the
minimum of energy for the excited states of $Be_{2}$ is reached at
shorter internuclear distance and is lower than that for the ground
state.


\section{Electronic excitations of a cluster in a unit cell}
\label{sec4}

\subsection{Computational method}

In \cite{Popov2008} was proposed an original technique for
describing the electronic structure of a cluster in a crystal when
the interaction of this cluster with the surrounding clusters is
very weak but cannot be ignored. Let us consider the main idea of
this method.

The total energy of electrons of a cluster in a unit cell can be
represented as the sum:
\begin{equation}
\label{33} E_{T}=E_{K}+E_{N}+E_{E}+ E_{C}+E_{X},
\end{equation}
which contains the kinetic energy of electrons
\begin{equation}
\label{34} E_{K}=\sum
\limits_{i,\textbf{k}}^{occ}\int\limits_{\Omega}
\Psi_{i\textbf{k}}^{*}(\textbf{r})\left(-\frac{\hbar^{2}}{2m}
\Delta\right)\Psi_{i\textbf{k}}(\textbf{r})d^{3}r,
\end{equation}
the potential energy of the electron–nucleus Coulomb interaction
\begin{equation}
\label{35} E_{N}=-\sum
\limits_{i,\textbf{k}}^{occ}\int\limits_{\Omega}\Psi_{i\textbf{k}}^{*}(\textbf{r})\sum
\limits_{p}\frac{Z_{i}e^{2}}{|\textbf{R}_{p}-\textbf{r}+\textbf{a}_{i}|}
\Psi_{i\textbf{k}}(\textbf{r})d^{3}r,
\end{equation}
the potential energy of the electron–electron Coulomb interaction
\begin{equation}
\label{36} E_{E}=\sum
\limits_{i,\textbf{k}}^{occ}\int\limits_{\Omega}\Psi_{i\textbf{k}}^{*}(\textbf{r})
\Psi_{i\textbf{k}}(\textbf{r}) \int\limits_{\Omega'}\sum \limits_{p}
\frac{\rho(\textbf{r}')e^{2}}
{|\textbf{R}_{p}-\textbf{r}+\textbf{r}'|}d^{3}r'd^{3}r,
\end{equation}
the potential energy of the nuclear Coulomb interaction
\begin{equation}
\label{37} E_{C}=\frac{1}{2}\sum
\limits_{i,j,p}\frac{Z_{i}Z_{j}e^{2}}
{|\textbf{R}_{p}-\textbf{a}_{i}-\textbf{a}_{j}|},
\end{equation}
and the energy of the electron exchange interaction
\begin{equation}
\label{38} E_{X}=e^{2}\sum \limits_{i,\textbf{k}}^{occ}\sum
\limits_{i',\textbf{k}'}^{occ}\int\limits_{\Omega}\int\limits_{\Omega'}
\Psi_{i\textbf{k}}^{*}(\textbf{r})\Psi_{i\textbf{k}}(\textbf{r}')
\Psi_{i'\textbf{k}'}^{*}(\textbf{r})\Psi_{i',\textbf{k}'}(\textbf{r})
G(\textbf{r}-\textbf{r}')d^{3}r'd^{3}r.
\end{equation}
Here,
\begin{equation}
\label{39} G(\textbf{r}-\textbf{r}')=-\sum
\limits_{p}\frac{exp\{i(\textbf{k}-\textbf{k}')\textbf{R}_{p}\}}
{|\textbf{R}_{p}-\textbf{r}+\textbf{r}'|}=-\frac{4\pi}{\Omega} \sum
\limits_{p}\frac{exp\{i(\textbf{K}_{p}+\textbf{k}-\textbf{k}')
(\textbf{r}-\textbf{r}')\}}
{|\textbf{K}_{p}+\textbf{k}-\textbf{k}'|^{2}}
\end{equation}
is the lattice Green’s function of the Laplace equation and

\begin{equation}
\label{40} \rho(\textbf{r})=\sum \limits_{i,\textbf{k}}^{occ}
|\Psi_{i\textbf{k}}(\textbf{r})|^{2}
\end{equation}
is the electron density in the unit cell. This electron density is
the sum of the wave functions squared over all filled states with
numbers $i$ and wave vectors $\textbf{k}$. The wave function in the
unit cell can be expanded in terms of the complete set of functions
$\{\varphi_{n}(\textbf{r})\}$; that is,
\begin{equation}
\label{41} \Psi_{i\textbf{k}}(\textbf{r})=\sum
\limits_{j,n}C_{ij}\left(\textbf{S}^{-1/2}\right)_{jn}\varphi_{n}(\textbf{r})
\end{equation}
with the overlap integral matrix $\textbf{S}$, which differs from a
unit matrix when the functions $\varphi_{n}(\textbf{r})$ are not
orthonormalized. Functions of the Gaussian type are chosen as the
functions $\varphi_{n}(\textbf{r})$ \cite{Huzinaga}. For the states
with symmetry $s$, the functions $\varphi_{n}(\textbf{r})$ are
represented in the form of the sum over the vectors $\textbf{R}_{p}$
of the direct lattice at large values $\alpha_{n}$, that is,
\begin{equation}
\label{42}
\varphi_{n}(\textbf{r})=\left(\frac{2\alpha_{n}}{\pi}\right)^{3/4}
\frac{1}{\sqrt{\Omega}}\sum
\limits_{p}exp\Big\{i\textbf{kR}_{p}-\alpha_{n}
|\textbf{R}_{p}-\textbf{r}+\textbf{a}_{n}|^{2}\Big\}, \qquad
\alpha_{n}\geq\alpha_{0},
\end{equation}
and the sum over the vectors $\textbf{K}_{p}$ of the reciprocal
lattice at small values $\alpha_{n}$, that is,
\begin{equation}
\label{43}
\varphi_{n}(\textbf{r})=\left(\frac{2\pi}{\alpha_{n}\Omega^{2}}\right)^{3/4}
\sum \limits_{p}exp \Big\{
i(\textbf{K}_{p}+\textbf{k})(\textbf{r}-\textbf{a}_{n})-
\frac{|\textbf{K}_{p}+\textbf{k}|^{2}}{4\alpha_{n}}\Big\}, \qquad
\alpha_{n}\leq\alpha_{0}.
\end{equation}
Here, $\textbf{a}_{n}$ is the vector specifying the position of the
atom in the unit cell of volume $\Omega$. It should be noted that
the constructed function (\ref{42}) satisfies the Bloch theorem and
that representation (\ref{43}) results from the transformation of
function (\ref{42}) into the sum over the reciprocal lattice
vectors. The functions with symmetries $p, d, f, ...$ can be easily
obtained from functions (\ref{42}) and (\ref{43}) with symmetry $s$
through the differentiation with respect to the components of the
vector $\textbf{a}_{n}$.

In order to improve the convergence of the summation over the
lattice in relationship (\ref{37}), we use the standard Ewald
method. The results obtained can be represented in the form
\begin{eqnarray}
\label{44} E_{C}=e^{2}\sum \limits_{i,j}Z_{i}Z_{j}\Big\{ \sum
\limits_{p}\frac{erfc(\sqrt{\alpha}|\textbf{R}_{p}-\textbf{a}_{i}+\textbf{a}_{j}|)}
{|\textbf{R}_{p}-\textbf{a}_{i}+\textbf{a}_{j}|} + \nonumber\\
+\frac{4\pi}{\Omega}\Big[\sum
\limits_{p\neq0}\frac{1}{\textbf{K}_{p}^{2}}
exp\Big(i\textbf{K}_{p}(\textbf{a}_{i}-\textbf{a}_{j})-
\frac{\textbf{K}_{p}^{2}}{4\alpha_{0}}\Big)-\frac{1}{4\alpha_{0}}\Big]\Big\},
\end{eqnarray}
where $erfc(x)$ is the complementary error function of the argument
$x$ and
\begin{equation}
\label{45} \alpha_{0}=\frac{K_{max}}{2R_{max}}
\end{equation}
is the parameter chosen from the condition for the identically rapid
convergence of the sum over the vectors $\textbf{R}_{p}$ of the
direct lattice to $R_{max}$ and the sum over the vectors
$\textbf{K}_{p}$ of the reciprocal lattice to $K_{max}$ in the expression (\ref{44}).

The iterative search for the self-consistent solution to the
algebraic eigenvalue problem can be considerably accelerated if we
determine the variational coefficients $C_{ij}$ averaged over all
values of the wave vector $\textbf{k}$ in the Brillouin zone. This
averaging can be performed in Eq.~(\ref{33}) when
calculating the sums over the wave vectors $\textbf{k}$ (the
integrals over the wave vector $\textbf{k}$ in the case of a
quasi-continuous distribution of states over the wave vectors) with
the use of the mean-value theorem. The narrower the $\textbf{k}$
band, i.e., the weaker the dependence of the wave functions on the
wave vector $\textbf{k}$, the smaller the error associated with the
averaging. The assumption that the wave functions and, hence, the
variational coefficients $C_{ij}$ depend weakly on the wave vector
$\textbf{k}$ is quite justified for the core states. This assumption
is also acceptable for sufficiently narrow completely filled bands.
The error introduced by the replacement of the integral over the
filled states of the wave vector $\textbf{k}$ by the integral over
the entire Brillouin zone can be substantially reduced by choosing a
larger unit cell in the lattice in order to decrease the sizes of
the Brillouin zone and to narrow the energy bands.

In order to solve the spectral problem and to determine the
variational coefficients $C_{ij}$ averaged over the Brillouin zone,
we consider Eq.~(\ref{33}) for the total energy. By using
expression (\ref{41}), we perform the summation in formulas
(\ref{34})-(\ref{36}) and (\ref{38}) over all values of the wave
vector $\textbf{k}$ (integration over the Brillouin zone) for each
energy band, which is hypothetically completely filled with
electrons. Then we use the variational principle. As a result, we
obtain the system of algebraic equations for calculating the
variational coefficients $C_{ij}$ averaged over the Brillouin zone.
The solution of the derived system of equations requires the
calculation of the matrix elements in the explicit form. For the
states with symmetry $s$, the matrix element of the overlap
integrals can be written in the form
\begin{equation}
\label{46}
S_{12}\equiv\int\varphi_{1}^{*}(\textbf{r})\varphi_{2}(\textbf{r})d^{3}r
d^{3}k=
\left(\frac{2\sqrt{\alpha_{1}\alpha_{2}}}{\alpha_{1}+\alpha_{2}}\right)^{3/2}
exp\left(-\frac{\alpha_{1}\alpha_{2}}{\alpha_{1}+\alpha_{2}}
|\textbf{a}_{1}-\textbf{a}_{2}|^{2}\right),
\end{equation}
the matrix element of the kinetic energy operator is given by the
formula
\begin{eqnarray}
\label{47} K_{12}\equiv\int
\varphi_{1}^{*}(\textbf{r})\left(-\frac{\hbar^{2}}{2m}\Delta\right)
\varphi_{2}(\textbf{r})d^{3}rd^{3}k=\nonumber\\
=\frac{\hbar^{2}}{2m}S_{12}\frac{\alpha_{1}\alpha_{2}}{\alpha_{1}+\alpha_{2}}
\left(6-\frac{4\alpha_{1}\alpha_{2}}{\alpha_{1}+\alpha_{2}}
|\textbf{a}_{1}-\textbf{a}_{2}|^{2}\right),
\end{eqnarray}
and the matrix element of the electron–nucleus interaction is
represented as
\begin{eqnarray}
\label{48} N_{12}\equiv-e^{2}\sum\limits_{i}Z_{i}\int
\varphi_{1}^{*}(\textbf{r})\varphi_{2}(\textbf{r})
\frac{d^{3}rd^{3}k}{|\textbf{R}_{p}-\textbf{r}+\textbf{a}_{i}|}=\nonumber\\
=-e^{2}S_{12}\sum\limits_{i}Z_{i}G^{(2)}(\alpha_{1}+\alpha_{2}).
\end{eqnarray}
Here, we have
\begin{equation}
\label{49} G^{(2)}(\alpha)=\frac{4\pi}{\Omega}
\Bigg[\sum\limits_{p\neq0}\frac{1}{K_{p}^{2}}
exp\Big(i\textbf{K}_{p}\textbf{R}-\frac{K_{p}^{2}}{4\alpha}\Big)
-\frac{1}{4\alpha}\Bigg],\qquad \alpha\leq\alpha_{0},
\end{equation}
\begin{equation}
\label{50}
G^{(2)}(\alpha)=G^{(2)}(\alpha_{0})+G^{(1)}(\alpha_{0})-G^{(1)}(\alpha),
\qquad \alpha\geq\alpha_{0},
\end{equation}
\begin{equation}
\label{51}
G^{(1)}(\alpha)\equiv\sum\limits_{p}\frac{erfc(\sqrt{\alpha}
|\textbf{R}_{p}+\textbf{R}|)}{|\textbf{R}_{p}+\textbf{R}|},
\end{equation}
\begin{equation}
\label{52}
\textbf{R}\equiv\frac{\alpha_{1}\textbf{a}_{1}+\alpha_{2}\textbf{a}_{2}}
{\alpha_{1}+\alpha_{2}}-\textbf{a}_{i}.
\end{equation}
The matrix element of the electron–electron Coulomb interaction with
symmetry $s$ is represented in the form
\begin{eqnarray}
\label{53} E_{1234}\equiv e^{2}\sum\limits_{p}\int
\varphi_{1}^{*}(\textbf{r})\varphi_{2}(\textbf{r})
\varphi_{3'}^{*}(\textbf{r}')\varphi_{4'}(\textbf{r}')
\frac{d^{3}r'd^{3}rd^{3}k'd^{3}k}{|\textbf{R}_{p}-\textbf{r}+\textbf{r}'|}
 =e^{2}S_{12}S_{34}G^{(2)}(\beta),
\end{eqnarray}
where the quantity $G^{(2)}(\beta)$ is calculated using the formulas identical to those used in the Ewald sums (\ref{49})-(\ref{51}) but for
\begin{equation}
\label{54}
\textbf{R}\equiv\frac{\alpha_{1}\textbf{a}_{1}+\alpha_{2}\textbf{a}_{2}}
{\alpha_{1}+\alpha_{2}}-
\frac{\alpha_{3}\textbf{a}_{3}+\alpha_{4}\textbf{a}_{4}}
{\alpha_{3}+\alpha_{4}},
\end{equation}
\begin{equation}
\label{55}
\beta\equiv\frac{(\alpha_{1}+\alpha_{2})(\alpha_{3}+\alpha_{4})}
{\alpha_{1}+\alpha_{2}+\alpha_{3}+\alpha_{4}}.
\end{equation}
The exchange energy matrix element calculated with the use of the
functions with symmetry $s$, i.\,e.
\begin{eqnarray}
\label{56} X_{1234}\equiv e^{2}\int
\varphi_{1}^{*}(\textbf{r})\varphi_{2}(\textbf{r}')\varphi_{3'}^{*}(\textbf{r})
\varphi_{4'}(\textbf{r}')G(\textbf{r}-\textbf{r}')d^{3}r'd^{3}rd^{3}k'd^{3}k=
\nonumber\\=
e^{2}\left(\frac{\pi^{2}}{(\alpha_{1}+\alpha_{4})(\alpha_{2}+\alpha_{3})}
\right)^{3/2}exp\left(-\frac{\alpha_{14}\alpha_{23}}{\alpha_{14}+\alpha_{23}}
|\textbf{a}_{14}-\textbf{a}_{23}|^{2}\right)G^{(3)}(\alpha_{14}+\alpha_{23}),
\end{eqnarray}
contains the factor $G^{(3)}(\alpha_{14}+\alpha_{23})$, which
possesses a satisfactory convergence of the sum over the vectors
$\textbf{R}_{p}$ of the direct lattice at large values
$(\alpha_{14}+\alpha_{23})\geq\alpha_{0}$ and a satisfactory
convergence of the sum over the vectors $\textbf{K}_{p}$ of the
reciprocal lattice at small values
$(\alpha_{14}+\alpha_{23})\leq\alpha_{0}$:
\begin{equation}
\label{57}
G^{(3)}(\alpha)=\sum\limits_{p}exp(-\alpha|\textbf{R}_{p}-\textbf{A}|^{2})
\frac{erf(\sqrt{\beta}|\textbf{R}_{p}\gamma+\textbf{B}|)}
{|\textbf{R}_{p}\gamma+\textbf{B}|}, \qquad \alpha\geq\alpha_{0},
\end{equation}
\begin{eqnarray}
\label{58}
G^{(3)}(\alpha)=\frac{1}{\Omega}\left(\frac{\pi}{\alpha}\right)^{3/2}
\sum\limits_{p}exp\left(i\textbf{K}_{p}\textbf{A}-\frac{K_{p}^{2}}{4\alpha}\right)
\times\nonumber\\ \times
erf\left(\sqrt{\frac{\alpha\beta}{\alpha+\beta}}\textbf{D}_{p}\right)
\frac{1}{\textbf{D}_{p}} , \qquad \alpha\leq\alpha_{0}.
\end{eqnarray}
Here, we used the following notation
\begin{eqnarray}
\label{59}
\alpha_{14}\equiv\frac{\alpha_{1}\alpha_{4}}{\alpha_{1}+\alpha_{4}};\qquad
\alpha_{23}\equiv\frac{\alpha_{2}\alpha_{3}}{\alpha_{2}+\alpha_{3}};\qquad
\alpha\equiv\alpha_{14}+\alpha_{23}; \nonumber\\
\beta\equiv\frac{(\alpha_{1}+\alpha_{2})(\alpha_{3}+\alpha_{4})}
{\alpha_{1}+\alpha_{2}+\alpha_{3}+\alpha_{4}};\qquad
\gamma\equiv\frac{\alpha_{4}}{\alpha_{1}+\alpha_{2}}-\frac{\alpha_{3}}
{\alpha_{2}+\alpha_{3}}; \nonumber\\
\textbf{A}\equiv\frac{\alpha_{14}\textbf{a}_{14}+\alpha_{23}\textbf{a}_{23}}
{\alpha_{14}+\alpha_{23}};\qquad
\textbf{a}_{14}=\textbf{a}_{1}-\textbf{a}_{4};\qquad
\textbf{a}_{23}=\textbf{a}_{2}-\textbf{a}_{3}; \\
\textbf{B}\equiv\frac{\alpha_{1}\textbf{a}_{1}+\alpha_{4}\textbf{a}_{4}}
{\alpha_{1}+\alpha_{4}}-\frac{\alpha_{2}\textbf{a}_{2}+\alpha_{3}\textbf{a}_{3}}
{\alpha_{2}+\alpha_{3}};\nonumber\\
\textbf{D}_{p}\equiv\textbf{B}+\gamma\textbf{A}+i\textbf{K}_{p}
\frac{\gamma}{2\alpha}.\nonumber
\end{eqnarray}

The total energy of the cluster in the unit cell derived from the aforementioned matrix elements in the following form:
\begin{equation}
\label{60}
E_{T}=\sum\limits_{n,m}(K_{nm}+N_{nm}+F_{nm})U_{nm}+E_{C},
\end{equation}
where
\begin{equation}
\label{61} U_{nm}=\sum\limits_{k}^{occ}\sum\limits_{i,j}C_{ki}C_{kj}
(\textbf{S}^{-1/2})_{in}(\textbf{S}^{-1/2})_{jm},
\end{equation}
\begin{equation}
\label{62}
F_{nm}=K_{nm}+N_{nm}+\sum\limits_{i,j}(2E_{nmij}-X_{nimj})U_{ij}.
\end{equation}

To describe cluster excitation in a lattice we used the same idea as formulated in section~\ref{3.1}. The use of this idea formally leads to additional contributions to matrix elements of the Fock operator (\ref{62}) of the matrix elements of the square angular momentum, see Eq.~(\ref{25}).

\begin{figure}[!hB]
\begin{minipage}[!hB]{0.45\linewidth}
\center{\includegraphics[width=0.9\linewidth]{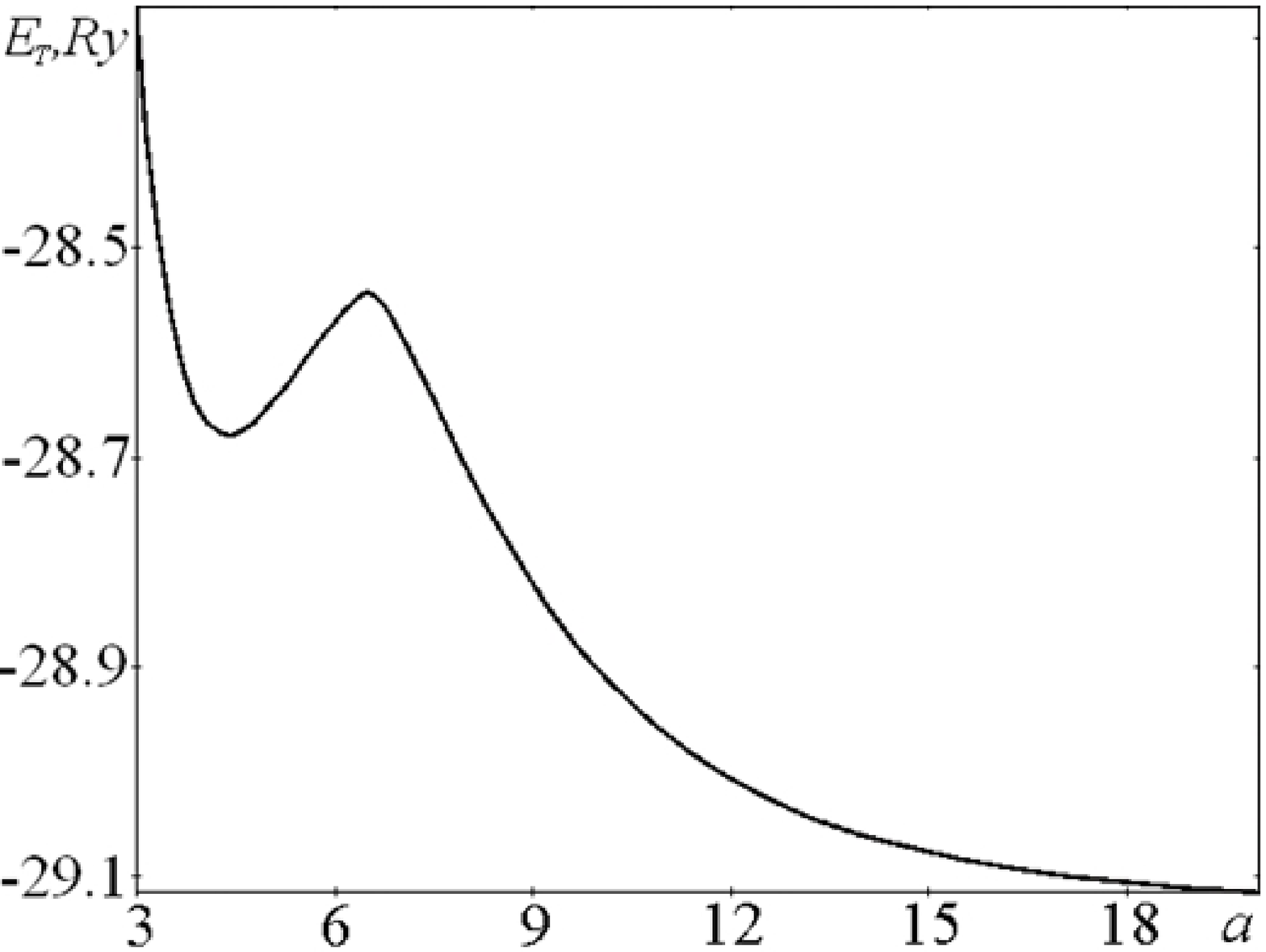}}
\caption{ Dependence of the total energy $E_{T}$ of electrons in
beryllium on the parameter $a$ of the hexagonal close-packed lattice
of the beryllium crystal.} \label{figure12}
\end{minipage}
\hfill
\begin{minipage}[!hB]{0.45\linewidth}
\center{\includegraphics[width=0.9\linewidth]{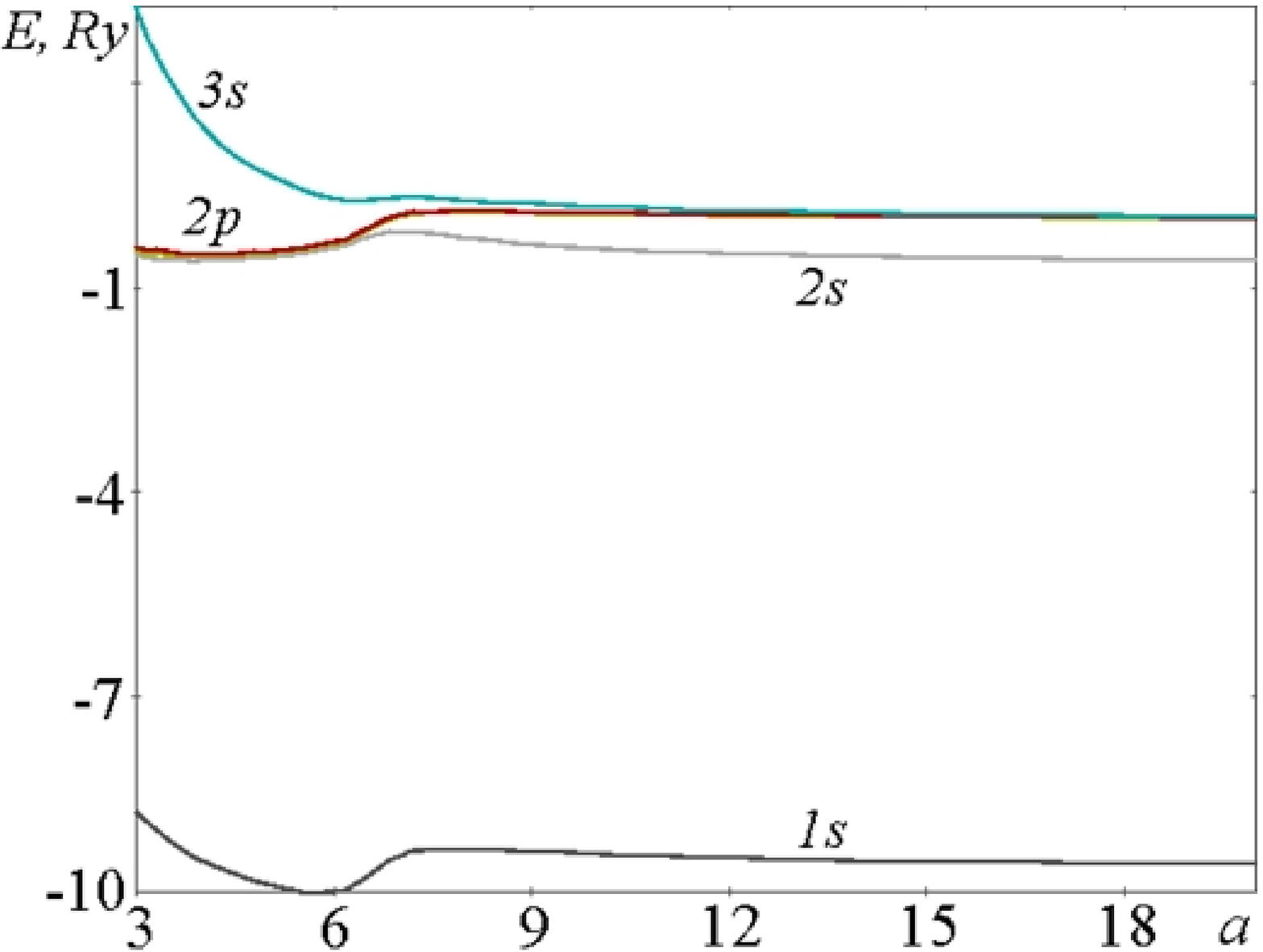}}
\caption{ Dependence of the energy $E$ of the electronic levels in
beryllium on the parameter $a$ of the hexagonal close-packed lattice
of the beryllium crystal.} \label{figure13}
\end{minipage}
\end{figure}

\subsection{Results and discussion}

All calculations were carried out for the hexagonal closed-packed beryllium crystal. The parameter $a$ was varied from twenty to three Bohr radii, and the parameter $c$ was calculated as $c=1.567a$, due to the fact that beryllium in the crystalline state has a hexagonal
closed-packed lattice with the ratio between of the parameters $c/a= 1.567$. It can be seen from Figure~\ref{figure12} that the total energy of the electronic system of the hexagonal closed-packed beryllium crystal is characterized by the minimum for $a=4.32$ bohr. This value lies within the limits of the spread of the experimental
data obtained by different authors. The energy states with
symmetries $1s, 2s, 2p,$ and $3s$ for the hexagonal closed-packed beryllium crystal are presented in Figure~\ref{figure13}. The lowest energy states with symmetries $1s$ and $2s$ are filled with electrons. The higher energy states with symmetries $2p$ and $3s$ are
not filled with electrons in the ground state for all parameters $a$ in the range from three to twenty bohr. The other unfilled states are not shown in Figure~\ref{figure13} for clarity. The results of
the calculations have demonstrated that, as the lattice parameter $a$ decreases, the $s–p$ hybridization of the states is observed beginning with seven bohr. It should be noted that, at these small parameters $a$, the mean $\textbf{k}$ band approximation used in the calculations can turn out to be invalid for beryllium. In our
opinion, the calculated parameter $a$ of the hexagonal closed-packed lattice of the beryllium crystal is in agreement with the experimental data, because this parameter was determined from the minimum of the total energy, which is an integrated characteristic.

\begin{figure}[!hB]
\begin{minipage}[!hB]{0.45\linewidth}
\center{\includegraphics[width=0.9\linewidth]{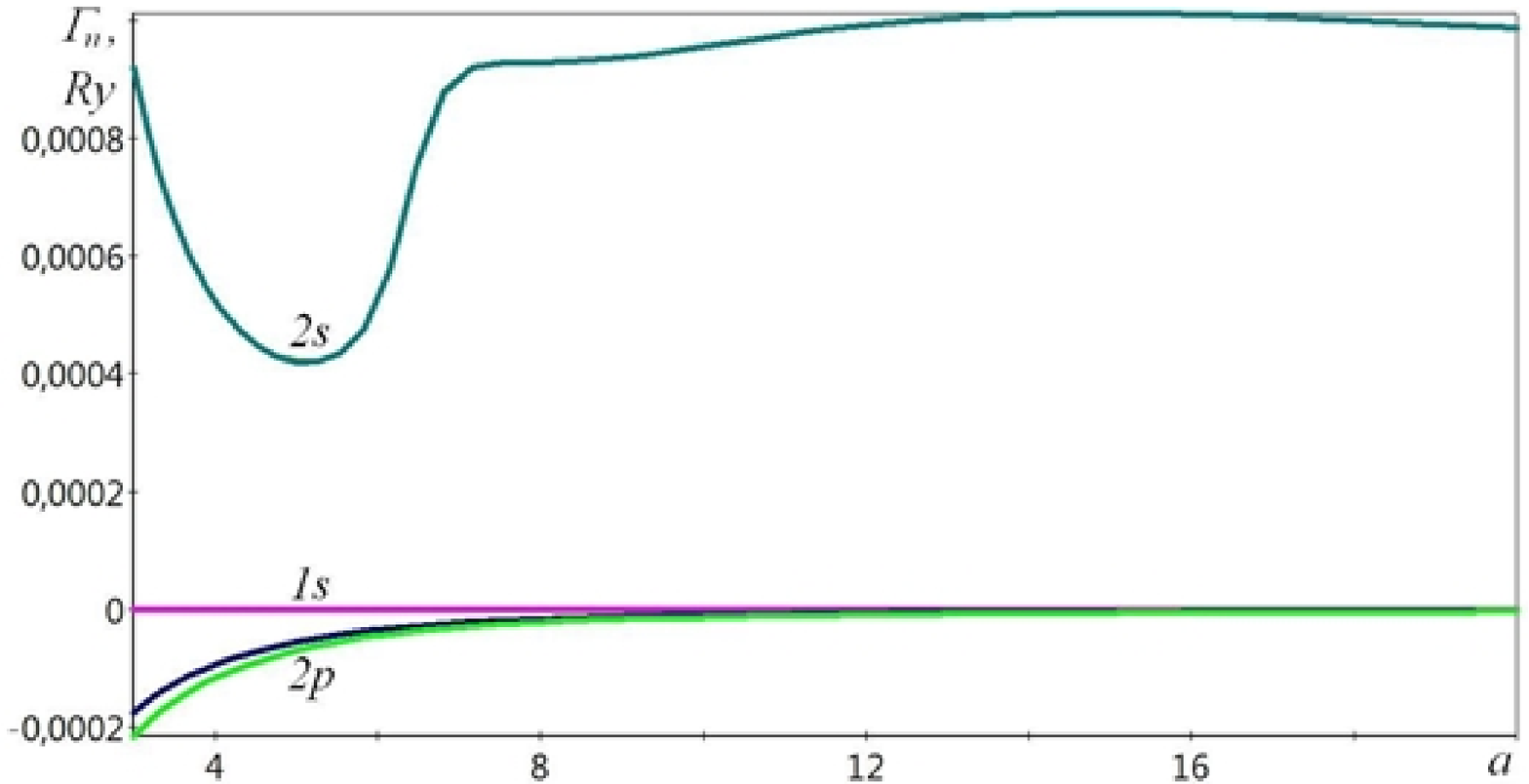}}
\caption{ Dependence of the imaginary part of the spectrum
$\Gamma_{n}$ of electrons in beryllium on the parameter $a$ of the
hexagonal close-packed lattice at $x=0$ and $y<4\cdot10^{-4}$.}
\label{figure14}
\end{minipage}
\hfill
\begin{minipage}[!hB]{0.45\linewidth}
\center{\includegraphics[width=0.9\linewidth]{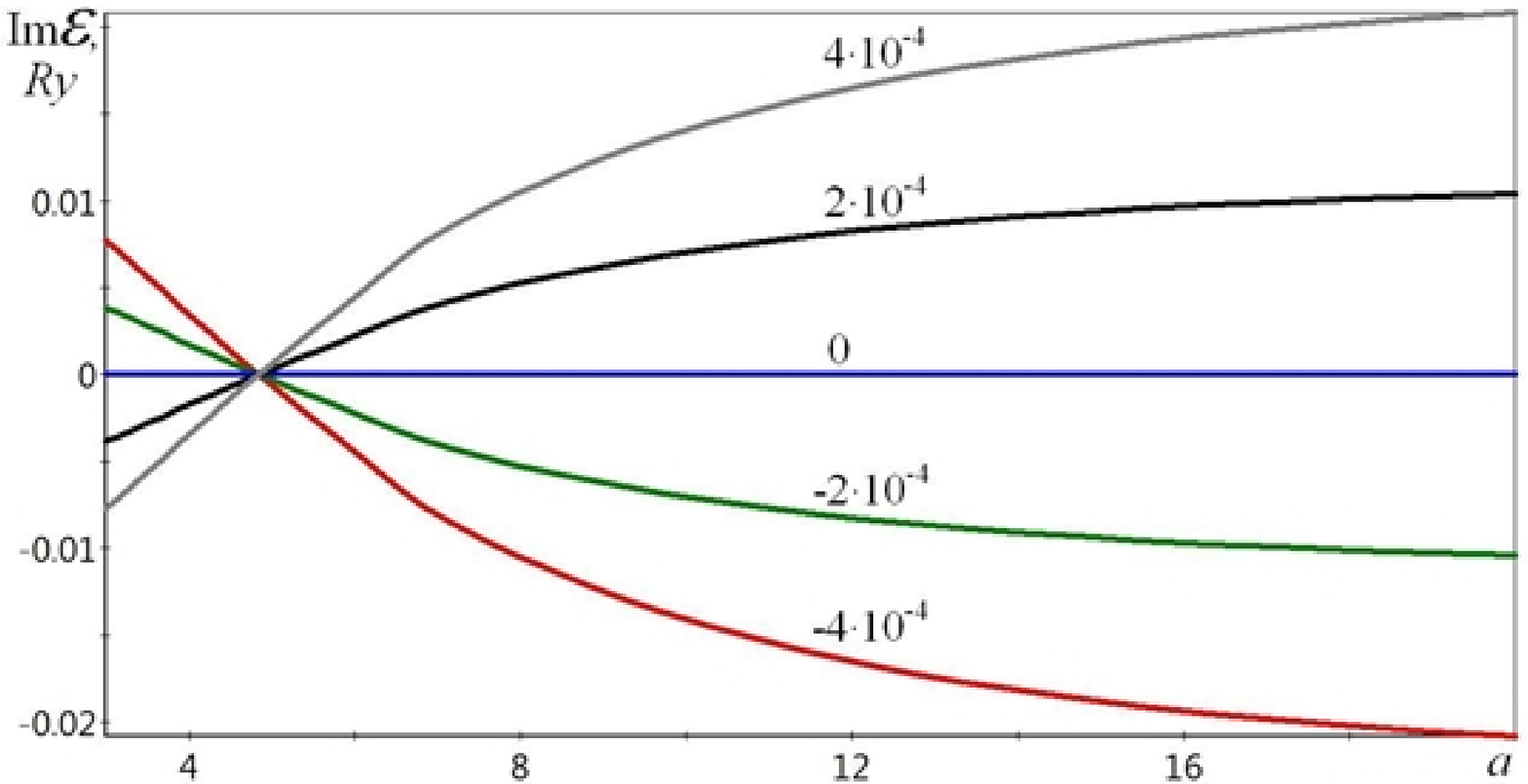}}
\caption{ Dependence of the imaginary part of the total energy
$Im\varepsilon$ of electrons in beryllium on the parameter $a$ of
the hexagonal close-packed lattice at $x=0$ and
$y=0,\pm2\cdot10^{-4},\pm4\cdot10^{-4}$.} \label{figure15}
\end{minipage}
\end{figure}

The presence of a high potential barrier of an order of 0.15 $Ry$ in
the hexagonal close-packed crystal lattice with the parameter $a$
approximately equal to seven bohr (Figure~\ref{figure12}) indicates
that the transition from the gaseous phase directly to the
crystalline phase (or in the backward direction) is extremely
hindered. The transition proceeding through a sequence of excited
metastable states with the formation of many-atom clusters is more
energetically favorable and, most likely, occurs in nature. This
assumption is confirmed by the calculation of the total energy of
equilibrium configurations of clusters in the hexagonal close-packed
crystal lattice with so large parameter that the interaction between
the clusters can be ignored. For this purpose, it will be sufficient, to take the lattice parameter equal to twenty Bohr radii.

The total energies of small-sized beryllium clusters per atom are
presented in Table~\ref{tab3}. This table presents the energies
for the most stable structures among all the structures revealed in
our investigations, i.e., for those characterized by the lowest
values of the total energy. It turns out that the existence of
$Be_{2}$ is energetically unfavorable. The existence of $Be_{3}$ is
also impossible, even though the energy of this cluster per atom is
considerably closer to the energy of an isolated atom $Be_{1}$. The
clusters $Be_{4}$, $Be_{5}$, and $Be_{6}$ are stable. It is quite
possible that crystallization occurs through the assembly of these (and, may be, still heavier) clusters. This statement is consistent with the data obtained in \cite{Srinivas,Sulka2013}, where the increasing of binding energy per atom with the size of the $Be$ cluster is established, and with the data reported in \cite{Wang}, where nonmetal-to-metal transitions in the $Be$ clusters were obtained with the increasing cluster size.

\begin{table}[!hB]
\caption{Total energies $E_{T}$ for equilibrium configurations of
small beryllium clusters} \label{tab3}
\begin{center}
\begin{tabular}{|c|c|c|c|c|c|c|}
\hline
Cluster &  $Be_{1}$ & $Be_{2}$ & $Be_{3}$ & $Be_{4}$ & $Be_{5}$ & $Be_{6}$ \\
\hline
$E_{tot}, Ry$ & -29.14 & -29.01 & -29.13 & -29.17 & -29.18 & -29.19 \\
\hline
\end{tabular}
\end{center}
\end{table}

Note that all lines shown in Figure~\ref{figure13} are practically the same for all values $|x|<5\cdot10^{-4}$ and $|y|<5\cdot10^{-4}$, in contrast to the imaginary part of the spectrum shown in Figure~\ref{figure14}. The imaginary part of the spectral line of $1s$-symmetry is equal to zero at all values, that corresponds to infinitely large value of the lifetime of electrons in that lowest energy state. The $2s$-state electrons have excitations with the
decay time, which is inversely proportional to the excitation energy, set by a parameter $y$.

Figure~\ref{figure15} shows the dependence of the imaginary part of
the total energy $Im\varepsilon$ of electrons in beryllium on the
parameter $a$ of the hexagonal close-packed lattice at $x=0$ and
$y=0,\pm2\cdot10^{-4},\pm4\cdot10^{-4}$. Note that $Im\varepsilon=0$
only near the equilibrium value of the lattice constant of the
hexagonal close-packed lattice for all values of $|x|<5\cdot10^{-4}$
and $|y|<5\cdot10^{-4}$. So, only this excitation is long-living.
All other states are quickly decay.



\section{Conclusion}

Using the proposed method, we have calculated the spectral
characteristics of atoms and small clusters of beryllium. A number of new phenomena, such as the energy collapse of an atom, anomalous splitting and mixing of energy states in beryllium atom is found.
Also the principal possibility of formation of
metastable excited states was shown, and the lifetime of these states was estimated. The predicted possibility of beryllium dimer formation at the distances between atoms near 4 and 8 bohr was experimentally
confirmed later. By using beryllium as an example, it is
demonstrated that the formation of a crystal lattice occurs via the formation of a set of small metastable clusters.

Thus, a considerable success has been achieved in atomistic simulations of electronic excitations of many-electron systems. The proposed methods for describing electronic excitations can give a complete view of the behavior of real matter, and also help to simulate the new properties of matter even not yet synthesized.








\end{document}